\documentclass[fleqn,usenatbib]{mnras}

\usepackage{graphicx}   
\usepackage{amsmath}    
\usepackage{amssymb}    

\usepackage{graphicx}
\usepackage{array}
\usepackage{amssymb}
\usepackage{float}
\usepackage{color}
\usepackage[utf8]{inputenc}  
\usepackage{amsmath}  
\usepackage{comment}
\usepackage{footnote}
\usepackage{longtable}
\usepackage[figuresright]{rotating}
\usepackage{booktabs}
\usepackage[T1]{fontenc}
\usepackage{footnote}
\usepackage{epstopdf}
\usepackage{hyperref}
\usepackage{cleveref}

\usepackage{newtxtext,newtxmath}

\title[Optical GeV connection in FSRQs]{\textbf{Correlation between optical and $\gamma$-ray flux variations in 
bright flat spectrum radio quasars}}

\author[Bhoomika et al.]{Bhoomika$^{1}$\thanks{E-mail: bhoomika@iiap.res.in},
C. S. Stalin$^{1}$,
S. Sahayanathan$^{2}$
\\
$^{1}$Indian Institute of Astrophysics, Block II, Koramangala, Bangalore 560034, India\\
$^{2}$Astrophysical Sciences Division, Bhabha Atomic Research Centre, Mumbai, India\\
}
\date{Last updated 2015 May 22; in original form 2013 September 5}

\pubyear{2015}

\begin{document}
\label{firstpage}
\pagerange{\pageref{firstpage}--\pageref{lastpage}}
\maketitle

\begin{abstract}
Blazars are known to show flux variations over a range of energies from low energy radio to high energy $\gamma$-rays. Cross-correlation analysis of the optical and $\gamma$-ray light curves in blazars shows that flux variations are generally correlated in both bands, however, there are exceptions. We explored this optical-GeV connection in four flat spectrum radio quasars (FSRQs) by a systematic investigation of their long term optical and $\gamma$-ray light curves. On analysis of the four sources, namely 3C 273, 3C 279, PKS 1510$-$089 and CTA 102 we noticed different behaviours between the optical and GeV flux variations. We found instances when (i) the optical and GeV flux variations are closely correlated (ii) there are optical flares without $\gamma$-ray counterparts and (iii) $\gamma$-ray flares without optical counterparts. To understand
these diverse behaviours, we carried out broad band spectral energy distribution (SED) modelling of the sources at different epochs using a one-zone leptonic emission model. The optical-UV emission is found to be dominated by emission from the accretion disk in the sources PKS 1510$-$089, CTA 102 and 3C 273, while in 3C 279, the synchrotron radiation from the jet dominates the optical-UV emission. Our SED analysis indicates that (i) correlated optical and $\gamma$-ray flux variations are caused by changes in the bulk Lorentz factor ($\Gamma$), (ii) $\gamma$-ray flares without optical counterparts are due to increase in $\Gamma$ and/or the electron energy density and (iii) an optical flare without $\gamma$-ray counterpart is due to increase in the magnetic field strength.
\end{abstract}


\begin{keywords}
galaxies: active - galaxies: nuclei - galaxies: jets - $\gamma$-rays:galaxies 
\end{keywords}




\section{Introduction}
Blazars, among the most luminous objects (10$^{42}$ $-$ 10$^{48}$ erg s$^{-1}$) in the Universe, are 
a class of active galactic nuclei (AGN) believed to be 
powered by accretion of matter onto super massive black holes
with masses greater than $\sim$ 10$^6$ M$_{\odot}$ situated at the centers of 
galaxies \citep{1969Natur.223..690L,1973A&A....24..337S}. These objects have relativistic jets
oriented close to the line of sight (within a few degrees) to the observer and
their radiation output is dominated
by non-thermal emission processes \citep{1993ARA&A..31..473A,1995PASP..107..803U}. 
They display rapid and large amplitude
flux variability over the entire accessible wavelength band  on a range of time scales from minutes to years \citep{1995ARA&A..33..163W,1997ARA&A..35..445U}. In addition to flux variations, blazars also show  high polarization \citep{1966ApJ...146..964K,1980ARA&A..18..321A} and polarization variability \citep{2005A&A...442...97A, 2010Natur.463..919A,2017ApJ...835..275R,2018ApJ...858...80R,2020MNRAS.492.1295P}. Blazars are
divided into flat spectrum radio quasars (FSRQs) and BL Lac objects
(BL Lacs) with FSRQs having strong emission lines in their optical/infra-red (IR) spectra
while BL Lacs have either featureless spectra or spectra with weak
emission lines, with equivalent widths $<$ 5 \AA. A physical
distinction between FSRQs and BL Lacs has been put forward
by \cite{2011MNRAS.414.2674G} with FSRQs having the 
ratio of the luminosity of the broad line region ($L_{BLR}$) to the 
Eddington luminosity ($L_{Edd}$) $>$ 5 $\times$ 10$^{-5}$.  A two
hump structure is evident in the broad band
spectral energy distribution (SED) of blazars,
the low energy component peaking in the IR-X-ray band and the high energy component
peaking in the MeV - GeV band \citep{1998MNRAS.299..433F,2016ApJS..224...26M}. In the leptonic scenario, the low energy component is attributed to synchrotron
emission process by the relativistic electrons in the jet, while the high
energy component is attributed to inverse Compton (IC) process \citep{2010ApJ...716...30A}.
The seed photons for IC emission could be the synchrotron photons from
the jet (synchroton self Compton SSC; \citealt{1981ApJ...243..700K,1985ApJ...298..114M, 
1989ApJ...340..181G}) as well as photons exterior to the jet (external compton EC;
\citealt{1987ApJ...322..650B}). These external
photons can be from the accretion disk \citep{1993ApJ...416..458D,1997A&A...324..395B}, 
the BLR \citep{1996MNRAS.280...67G,1994ApJ...421..153S} and the 
torus \citep{2000ApJ...545..107B,2008MNRAS.387.1669G}.

\begin{table*}
\label{table-1}
\caption{Details of the objects analysed in this work. The mean $\gamma$-ray flux in 
the 100 MeV - 300 GeV band is in units of $10^{-7}$ ph $cm^{-2}$ $s^{-1}$ and $\Gamma_{p}$ is the $\gamma$-ray photon index in the 100 MeV - 300 GeV band. The values of $\Gamma_{p}$ are from \citep{2015ApJ...810...14A}. }
\begin{tabular} {lccrccr} \hline
Name           & 4FGL name        & $\alpha_{2000}$ & $\delta_{2000}$  & $z$ & $\Gamma_{p}$ & $\gamma$-ray flux \\ \hline
PKS 1510$-$089 &  4FGL J1512.8$-$0906 & 15:12:50.53 & $-$09:05:59.83 &  0.360 & 2.364 & 9.13 \\
3C 273         &  4FGL J1229.1+0202   & 12:29:06.70 &   +02:03:08.60 &  0.158 & 2.661 & 6.73 \\
3C 279         &  4FGL J1256.1$-$0547 & 12:56:11.17 & $-$05:47:21.52 &  0.536 & 2.343 & 8.79 \\
CTA 102        &  4FGL J2232.5+1143   & 22:32:36.41 &   +11:43:50.90 &  1.037 & 2.520 & 13.70 \\ \hline
\end{tabular}
\end{table*}

The observed broad band SED of blazars are generally explained satisfactorily by
leptonic models, however, there are exceptions, wherein the observed
SED is interpreted by hadronic \citep{2003APh....18..593M,2013ApJ...768...54B,
2019MNRAS.486.1781R} or lepto-hadronic models \citep{2016ApJ...826...54D,2016ApJ...817...61P}.
In the hadronic models of blazars, the high energy emission is 
due to synchrotron emission from protons that are accelerated to 
relativistic energies \citep{2003APh....18..593M,2000NewA....5..377A} or from pair cascades 
initiated by proton-proton or proton-photon interactions
\citep{1993A&A...269...67M}. Recent observations indicate that 
a single model might be inadequate to explain the SED of a source at all times.
For example, in the broad band SED analysis of 3C 279 at various epochs, it has been found that 
leptonic model explains the SED during the March - April 2014 flare \citep{2015ApJ...803...15P}, while the 
SED during the flare in December 2013 is better fit by lepto-hardronic models 
\citep{2016ApJ...817...61P}. TXS 0506+056 is the first blazar associated with the detection of
neutrinos by the IceCube neutrino observatory on 22 September 2017 and this 
was coincident in direction and time with a $\gamma$-ray flare from TXS 0506+056. 
This gives observational evidence of hadronic emission in blazars \citep{2018Sci...361..147I}.
Also, recently another blazar has been found to be spatially coincident with 
the IceCube neutrino event IC-200107A \citep{2020arXiv200306012P}.
Thus, it is very clear that we do not yet fully understand the emission processes 
that contribute to the high energy emission in blazars.

Broad band SED modelling of blazars is often used to constrain the 
hadronic v/s leptonic scenario for the production of high energy $\gamma$-ray 
emission in them. An alternative to this SED based approach is the one based
on carrying out a comparative analysis of the flux variations in the optical and
$\gamma$-ray bands. In the leptonic scenario, as the relativistic electrons in
the jets of blazars are responsible for both the optical and $\gamma$-ray emission
a close correlation is expected between the optical and $\gamma$-ray flux variations
\citep{2007Ap&SS.309...95B}. Alternatively, in the hadronic model of emission from 
blazars, though the optical emission is dominated by electron synchrotron, the
$\gamma$-ray emission could be from proton synchrotron, and therefore, 
a correlation between the optical and $\gamma$-ray flux variations may not be 
expected \citep{2001APh....15..121M}. Thus, by a systematic investigation of the correlation
between the optical and $\gamma$-ray flux variations in a sample of blazars, it would
be possible to constrain the leptonic v/s hardonic emission from blazar jets.
An alternative to SED modelling and optical - $\gamma$-ray studies to distinguish
between the leptonic and hadronic scenarios in the high energy emission from 
FSRQs is through their X-ray polarization. According to \cite{2013ApJ...774...18Z}
X-ray polarization in blazars will be different in these two scenarios. X-ray polarimetric
observations in the future from the Imaging X-ray Polarimetry Explorer (IXPE; 
\citealt{2016SPIE.9905E..17W}) will be able to constrain the origin of high 
energy emission in blazars.

The launch of the {\it Fermi} Gamma-ray Space Telescope (hereinafter {\it Fermi}; \citealt{2009ApJ...697.1071A}) in the year 2008 has enabled
investigation of the long term $\gamma$-ray flux variability characteristics
of blazars \citep{2020A&A...634A..80R} that dominate the extragalactic $\gamma$-ray sky. Prior to {\it Fermi}, the 
availability of long term $\gamma$-ray light curves of blazars were limited. However,
today we know about 3000 blazars that are detected by {\it Fermi} \citep{2019arXiv190510771T} and
most of them have $\gamma$-ray light curves spanning more than 10 
years\footnote{https://fermi.gsfc.nasa.gov/ssc/data/access/lat/msl\_lc/} suitable
for long term $\gamma$-ray variability studies. In support of {\it Fermi}, 
ground based monitoring observations in the optical and IR are being carried
out by the Small and Moderate Aperture Research Telescope System 
(SMARTS\footnote{http://www.astro.yale.edu/smarts/glast/home.php}; \citealt{2009ApJ...697L..81B}) 
and the Steward Observatory\footnote{http://james.as.arizona.edu/~psmith/Fermi/\#mark6} \citep{2009arXiv0912.3621S}.  
These observations in the optical and IR bands serve as a valuable data set to 
study the correlations between the optical and $\gamma$-ray flux variations in 
blazars. Studies carried out on these lines have led to varied results. 
Few studies demonstrated that the $\gamma$-ray flares in blazars are correlated
with optical flares with or without lag \citep{2009ApJ...697L..81B,2012ApJ...749..191C,2014ApJ...783...83L,2015MNRAS.450.2677C}. However, studies of this kind carried out on more objects have found that the 
optical and $\gamma$-ray flux variations are not correlated all times and there are objects where
$\gamma$-ray flares are detected without an optical counterpart \citep{2011ApJ...736L..38V,2013ApJ...779..174D,2015ApJ...804..111M}.
Similarly, prominent optical flares with no corresponding $\gamma$-ray flares are also known in some
objects \citep{2013ApJ...763L..11C, 2014ApJ...797..137C,2019MNRAS.486.1781R}. Recently, \cite{2019ApJ...880...32L}
looked for the presence/absence of correlated optical and $\gamma$-ray flux variations in a large
sample of {\it Fermi} blazars. To further probe the prevalence of
anomalous optical and $\gamma$-ray flux variability in blazars and understand their physical characteristics through broad band SED modelling, we carried
out a systematic analysis of the $\gamma$-ray flux variability of blazars that
have optical and IR monitoring data available in the archives. Here, we present our results
on four FSRQs. Results on the BL Lacs analyzed as part of this investigation will be presented elsewhere. In section 2, we provide the details on the selection of the objects for 
this program. The data used in this work is explained in Section 3 followed by the analysis in 
Section 4. The results are presented in Section 5 followed by the summary in the final section.

\begin{table*}
\caption{Details of the epochs considered for detailed light curve analysis, 
SED modelling and spectral analysis. The $\gamma$-ray fluxes in the 100 MeV to 300 GeV band 
are in units of 10$^{-6}$ ph cm$^{-2}$ s$^{-1}$ and the optical fluxes in the V-band are in units of 
10$^{-11}$ erg cm$^{-2}$ s$^{-1}$}
\begin{tabular} {lcccccccl} \hline
              &      & \multicolumn{2}{c}{MJD}  & \multicolumn{2}{c}{Calendar date}  & \multicolumn{2}{c}{Mean flux} & \\
Name          & ID   &  Start   & End            & Start          & End                & $\gamma$      & Optical  & Remark  \\ \hline    
PKS 1510$-$089  &  A   &  54937   & 54957         & 16-04-2009     & 06-05-2009        & 2.97               & 1.07 & $\gamma$-ray flare with no optical flare  \\
              &  B   &  54951   & 54971         & 30-04-2009     & 20-05-2009        & 2.26               & 1.94 & $\gamma$-ray flare and optical flare  \\
              &  C   &  55757   & 55777         & 15-07-2011     & 04-08-2011        & 1.10               & 0.66 & $\gamma$-ray flare with no optical flare  \\
              &  D   &  56062   & 56162         & 15-05-2012     & 23-08-2012        & 0.44               & 0.62 & Quiescent state  \\
              &  E   &  57105   & 57125         & 24-03-2015     & 13-04-2015        & 3.12               & 0.95 & $\gamma$-ray flare with no optical flare  \\ 
              &  F   &  57157   & 57177         & 15-05-2015     & 04-06-2015        & 3.17               & 2.30 & $\gamma$-ray flare and optical flare \\  \hline
3C 273        &  A   &  55265   & 55285         & 10-03-2010     & 30-03-2010        & 1.53               & 16.9 & $\gamma$-ray flare with no optical flare \\ 
              &  B   &  56450   & 56550         & 07-06-2013     & 15-09-2013        & 0.40               & 16.7 &  Quiescent state  \\ \hline
3C 279        &  A   &  55290   & 55390         & 04-04-2010     & 13-07-2010        & 0.26               & 0.14 & Quiescent state   \\
              &  B   &  56742   & 56762         & 26-03-2014     & 15-04-2014        & 2.21               & 2.15 & $\gamma$-ray flare with no optical flare  \\ 
              &  C   &  57178   & 57198         & 05-06-2015     & 25-06-2015        & 3.94               & 1.42 & $\gamma$-ray flare with no optical flare \\ 
              &  D   &  57828   & 57848         & 16-03-2017     & 05-04-2017        & 2.33               & 4.25 & optical flare but no $\gamma$-ray flare \\ \hline
CTA 102       &  A   &  55840   & 55940         & 06-10-2011     & 14-01-2012        & 0.31               & 0.39 & Quiescent state \\ 
              &  B   &  57740   & 57750         & 18-12-2016     & 28-12-2016        & 10.3               & 44.5 & $\gamma$-ray flare and optical flare  \\ \hline
\end{tabular}
\label{table-2}
\end{table*}

\section {Sample}
For this work, we first selected all sources that are classified as FSRQs in 
the third catalog of AGN detected by the large area telescope (LAT) onboard 
{\it Fermi} (3LAC; \citealt{2015ApJ...810...14A}). For the selected 
FSRQs we then looked into their one day binned $\gamma$-ray light curves given at the {\it Fermi} site \footnote{https://fermi.gsfc.nasa.gov/ssc/data/access/lat/msl\_lc/} and 
selected those sources that have at least one flare with the $\gamma$-ray flux 
exceeding 10$^{-6}$ photons cm$^{-2}$ s$^{-1}$. This lead us to a sample of 84 sources. For those 84 sources, we looked at the archives of SMARTS for the availability of optical and IR data overlapping the duration of $\gamma$-ray light curves. For 40 out of the 84 sources we found data in SMARTS. Of these 40, three sources namely 3C 454.3, PKS 1510$-$089 and 3C 279 have the largest number of data points in the optical and IR bands with the total exceeding 500. To these three, we added two more sources namely CTA 102 and 3C 273 due to their high $\gamma$-ray activity states \citep{2016ATel.9869....1C,2009ATel.2168....1B}. Thus, our final sample for correlated optical - GeV studies consists of five sources. Of these five, results for one source 3C 454.3 is already published in \cite{2019MNRAS.486.1781R}. In this work we present our results on the analysis of the remaining four sources. The details of these four sources are given in Table \ref{table-1}. A brief description about them are given below:
\subsection{PKS 1510$-$089} 
It was identified as a quasar firstly by \cite{1966AuJPh..19..559B}  
with a visual magnitude of 16.5 mag. It is one of the most variable FSRQs in the 
3FGL catalog. Located at a 
redshift of $z$ = 0.361 \citep{1996AJ....112...62T}, it is powered by a black hole of mass 
5.4 $\times$ 10$^8$ M$_{\odot}$ and accretes  at the rate of
0.5 M$_{\odot}$/year \citep{2010ApJ...721.1425A}.  It has been detected at very high energies by HESS \citep{2013A&A...554A.107H} and 
MAGIC (Major Atmospheric Gamma-Ray Imaging Cherenkov; \citealt{2014A&A...569A..46A}). 
This source has been studied for multi-wavelength flux variability \citep{2017ApJ...844...62P,2013MNRAS.430.1324N} as well as subjected to few SED modelling
studies \citep{2019ApJ...883..137P,2012ApJ...760...69N}. Considering radio observations with the VLBA coupled with optical long term 
monitoring data \cite{2005MNRAS.361..155W} argued for the presence of a binary black hole in
PKS 1510$-$089. 

\subsection{3C 273} 
3C 273, the first quasar discovered 
by \cite{1963Natur.197.1040S} at a redshift $z$ = 0.158 has a large scale radio jet with 
a projected size of 57 kpc \citep{2006ARA&A..44..463H}. It was the first 
quasar that was discovered in the $\gamma$-ray band in the energy range of 50$-$500 MeV 
\citep{1978Natur.275..298S}. It was later detected 
by the Energetic Gamma- Ray  Experiment Telescope (EGRET; \citealt{1999ApJS..123...79H}) and then by {\it Fermi}. It has been
studied for flux variations in the optical \citep{2017ApJS..229...21X} and 
also has been found to show dramatic 
variations in the $\gamma$-ray band from {\it Fermi} observations
\citep{2010ApJ...714L..73A}. The $\gamma$-ray outburst in 2009 was 
explained by a time dependent one zone synchrotron self-Compton model 
\citep{2013MNRAS.431.2356Z}. 

\subsection{3C 279} At a  redshift of $z$ = 0.536 \citep{1965ApJ...142.1667L}, 3C 279 was among the blazars 
that were discovered as emitters of $\gamma$-rays  by EGRET \citep{1992ApJ...385L...1H}. 
In the GeV-TeV range it was first detected by the ground based atmospheric 
Cherenkov  experiment MAGIC \citep{2008Sci...320.1752M}. It has been recently suggested that
3C 279 hosts a supermassive black hole binary at its center \citep{2019A&A...621A..11Q}.
The source is found to show flux variations over a range of wavelengths such as
radio \citep{1966ApJ...146..634P}, optical \citep{1967ApJ...147..901O} and 
$\gamma$-rays \citep{1992ApJ...385L...1H}. It has also been studied for correlated
variations over different wavebands \citep{2008ApJ...689...79C}. {\it Fermi} observations
have revealed minute scale flare in this source with a shortest flux
doubling time scale lesser than 5 minutes during the outburst in 2015 \citep{2017AIPC.1792e0015H}. 
In addition to flux variability studies, it has also been studied via broad band SED modelling during
various activity states. The flares at different epochs of the source
were explained by leptonic process \citep{2015ApJ...803...15P,2019MNRAS.484.3168S}, lepto-hardonic process 
\citep{2018ApJ...863...98P}  as well as  hadronic processes \citep{2017MNRAS.467L..16P}.
These observations and subsequent modelling clearly indicate that the same emission mechanisms
are not responsible for the high energy emission we receive from the source
at all times.

\subsection{CTA 102} This FSRQ at a redshift of $z$ = 1.037 \citep{1965ApJ...141.1295S}
is highly polarized \citep{1981ApJ...243...60M} and variable in the optical band \citep{1986ApJ...310..325M}.  
It was detected in the $\gamma$-ray band both by EGRET \citep{1994ApJS...94..551F}  
and {\it Fermi} \citep{2009ApJ...700..597A}. It has been studied for flux variations
across different wavebands \citep{2018A&A...617A..59K} and minute like time 
scales of variability were detected in the optical \citep{2009AJ....138.1902O}  
and $\gamma$-ray bands \citep{2018ApJ...854L..26S}.

\begin{figure*}
\hspace*{-3cm}
\includegraphics[width=1.3\textwidth]{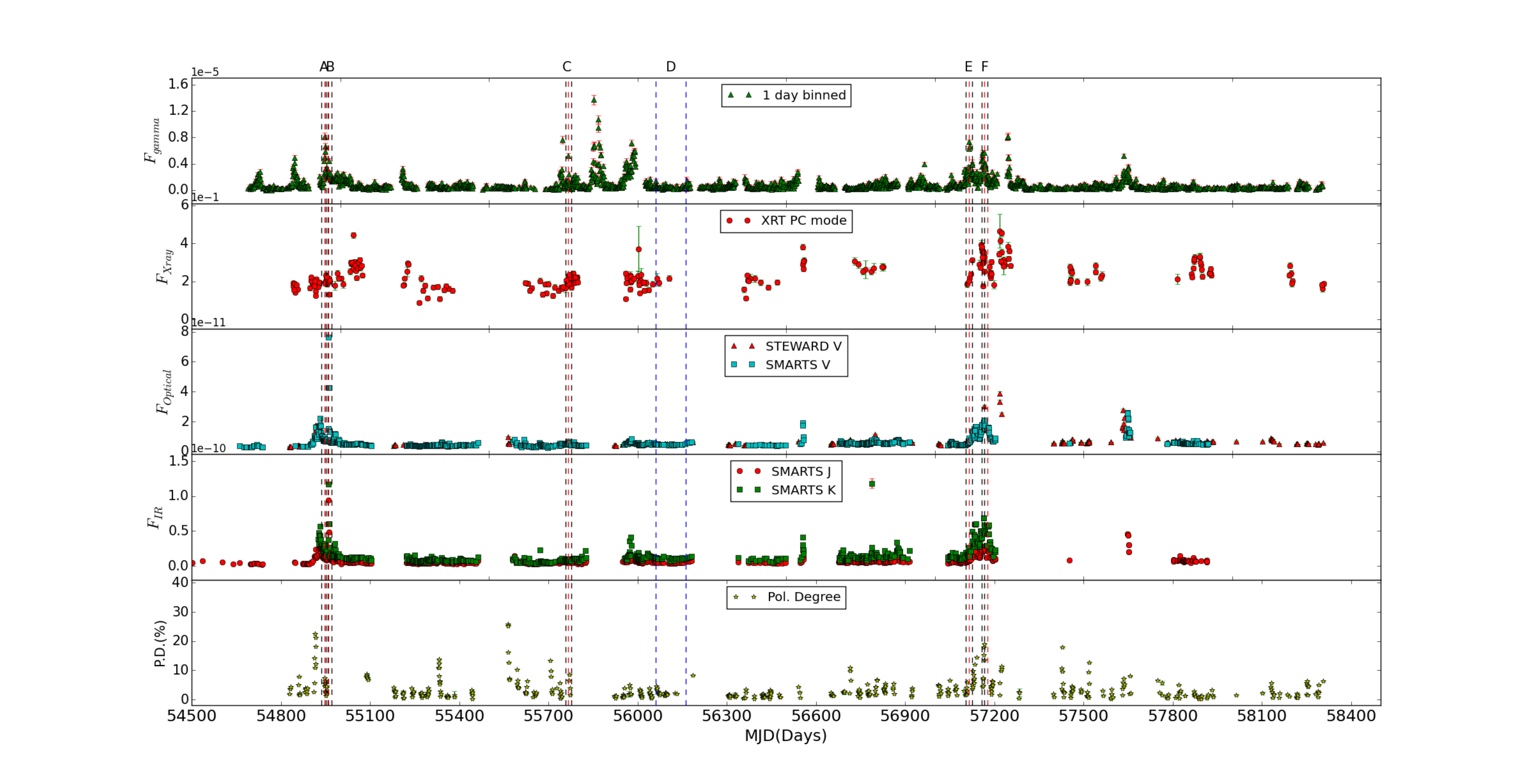}
\caption{Light curves of the source PKS 1510$-$089. The top one is the one day 
binned $\gamma$-ray light curve (in units of $10^{-5}$ ph $cm^{-2}$ $s^{-1}$), the second panel from the top is the X-ray 
light curve (in units of counts/sec), the next two panels are the optical (in units of $10^{-11}$ erg $cm^{-2}$ $s^{-1}$) and the IR (in units of $10^{-10}$ erg $cm^{-2}$ $s^{-1}$) light curves and the bottom panel is the optical V-band polarization. The peak of either the
optical or $\gamma$-ray light curve is shown by red dotted lines, while the
two black solid lines on either side of the red line correspond to a width of 10
days each. The two blue lines show the quiescent period of 100 days. For the
$\gamma$-ray light curve, upper limits are not shown and only points with TS $>$ 9
are plotted.}
\label{figure-1}
\end{figure*}  

\begin{figure*}
\vbox{
\hbox{
\includegraphics[scale=0.45]{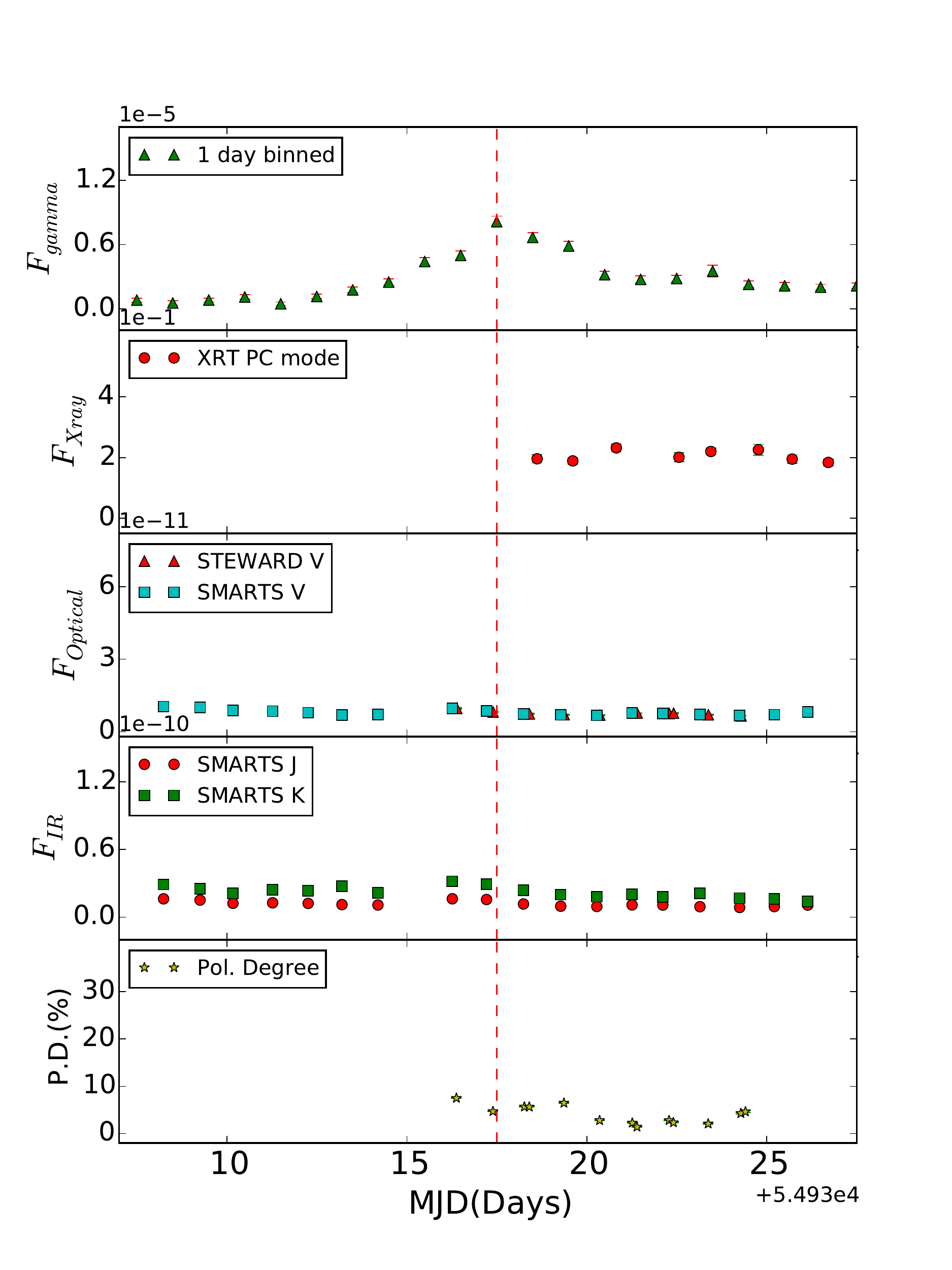}
\includegraphics[scale=0.45]{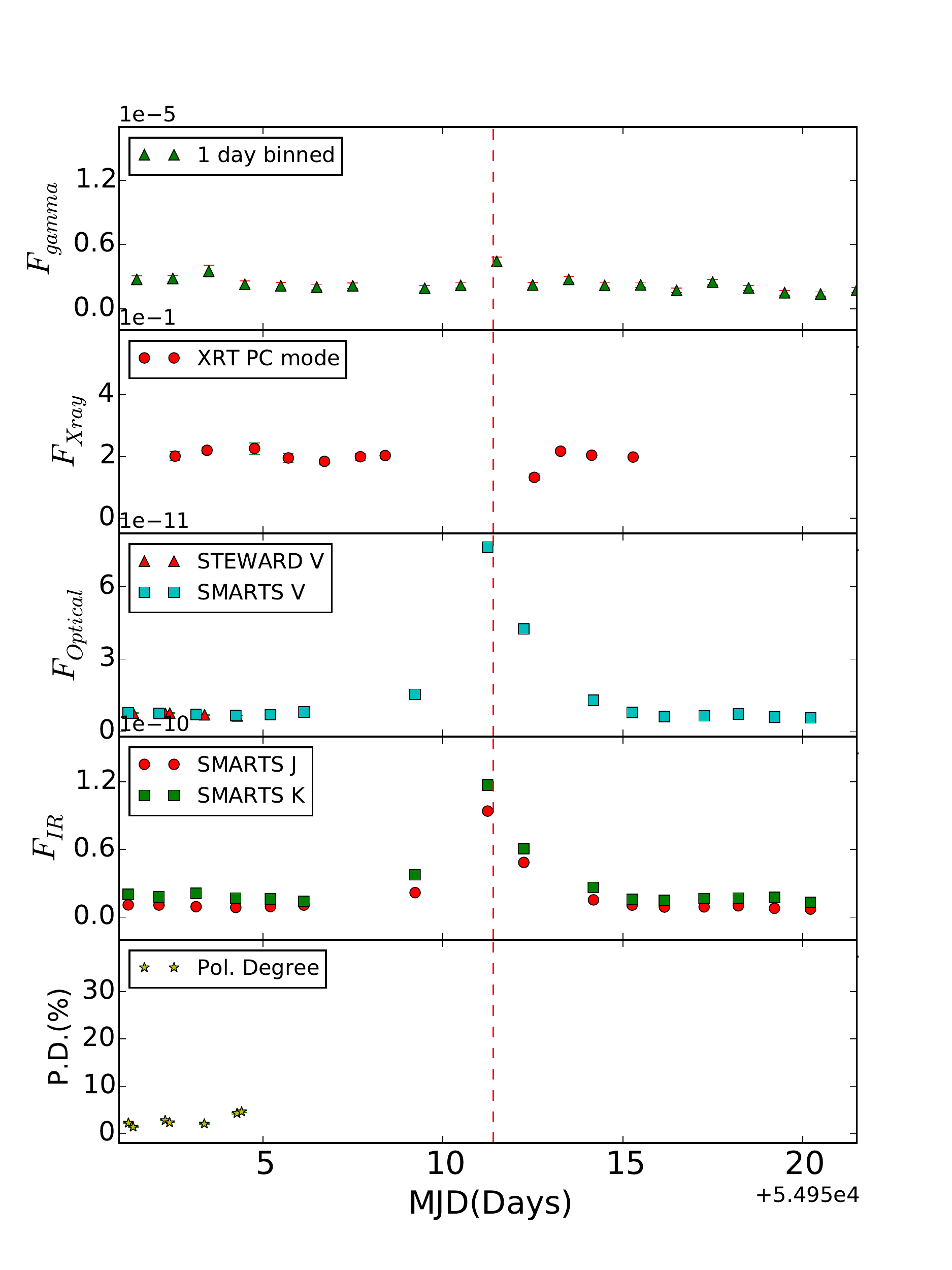}
     }
\hbox{
\includegraphics[scale=0.45]{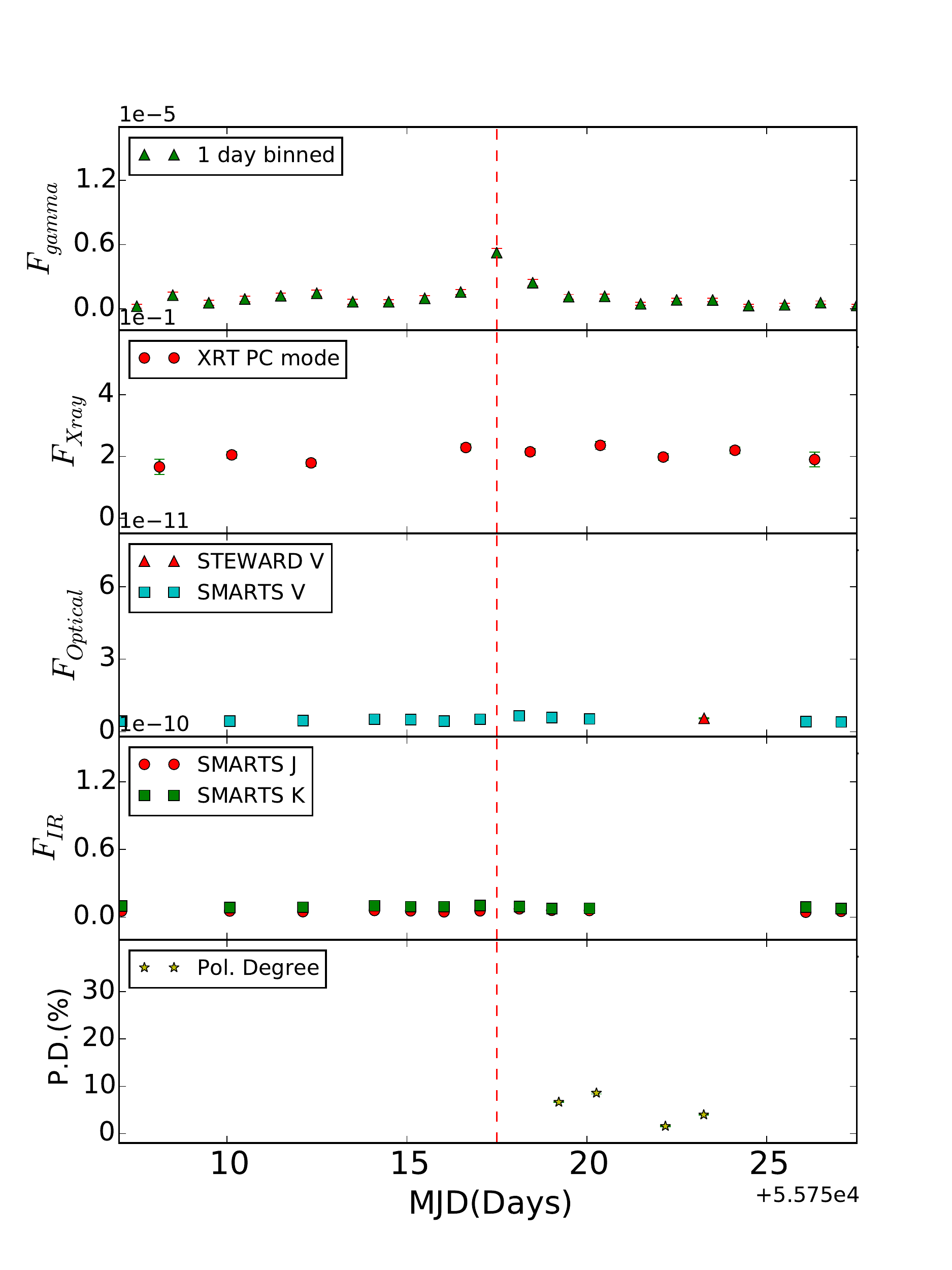}
\includegraphics[scale=0.45]{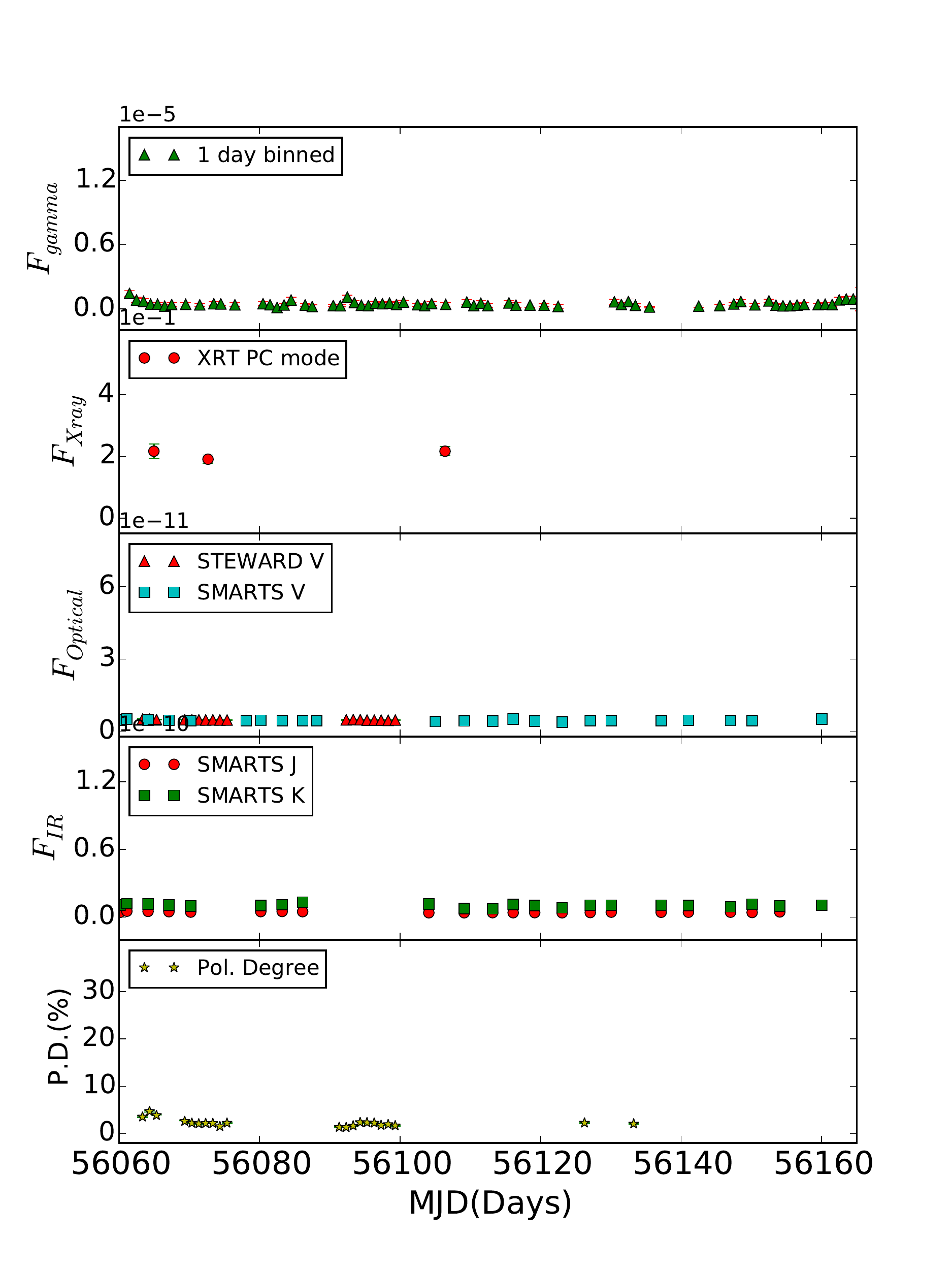}
    }
}
\caption{Multi-wavelength light curves for epochs A (top left), B (top right), 
C (bottom left) and D (bottom right) for the source PKS 1510$-$089.  In all the 
panels $\gamma$-ray fluxes are in units of 10$^{-5}$ ph cm$^{-2}$ s$^{-1}$. The 
optical fluxes are in units of 10$^{-11}$ erg cm$^{-2}$ s$^{-1}$ and the IR fluxes are 
in the units of 10$^{-10}$ erg cm$^{-2}$ s$^{-1}$. The vertical dotted line 
shows the peak of the optical/$\gamma$-ray flare.}
\label{figure-2}
\end{figure*}

\begin{figure*}
\hbox{
\includegraphics[scale=0.45]{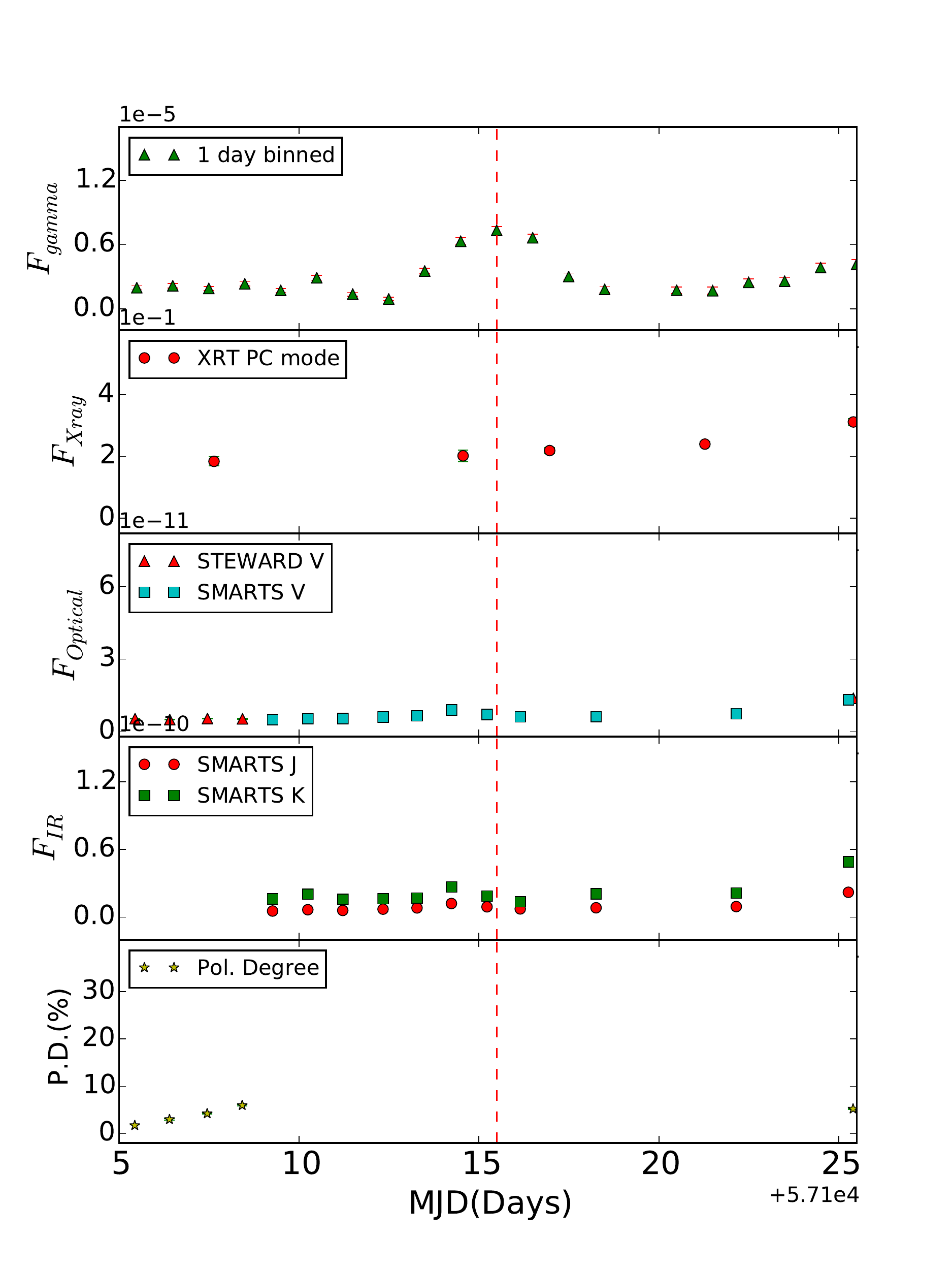}
\includegraphics[scale=0.45]{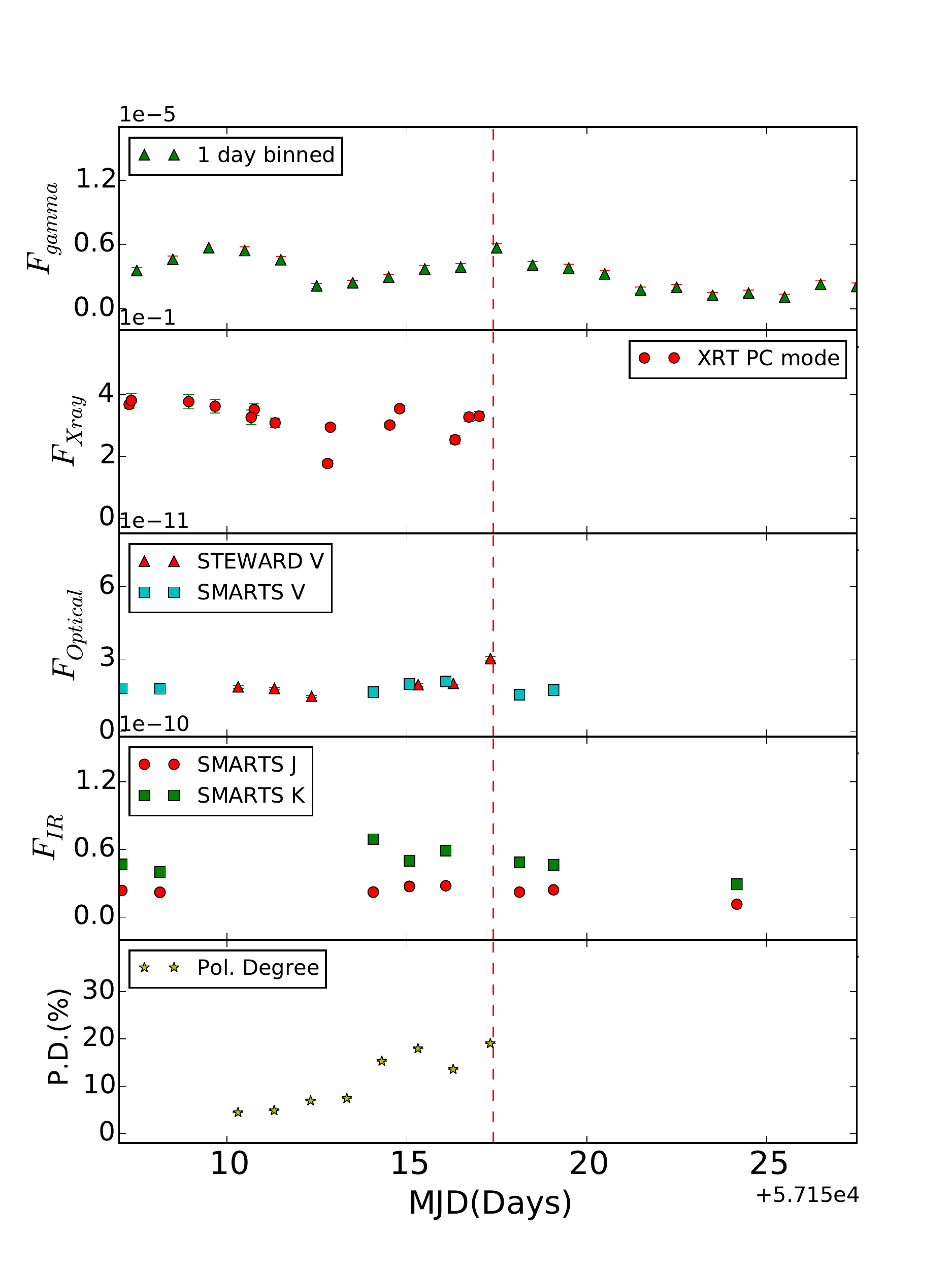}
     }
\caption{Muti-wavelength light curves for the source PKS 1510$-$089 during epochs
E (left) and F (right). Symbols and lines are as in Fig. \ref{figure-2}.}
\label{figure-3}
\end{figure*}

\begin{figure*}
\hspace*{-3cm}
\includegraphics[width=1.3\textwidth]{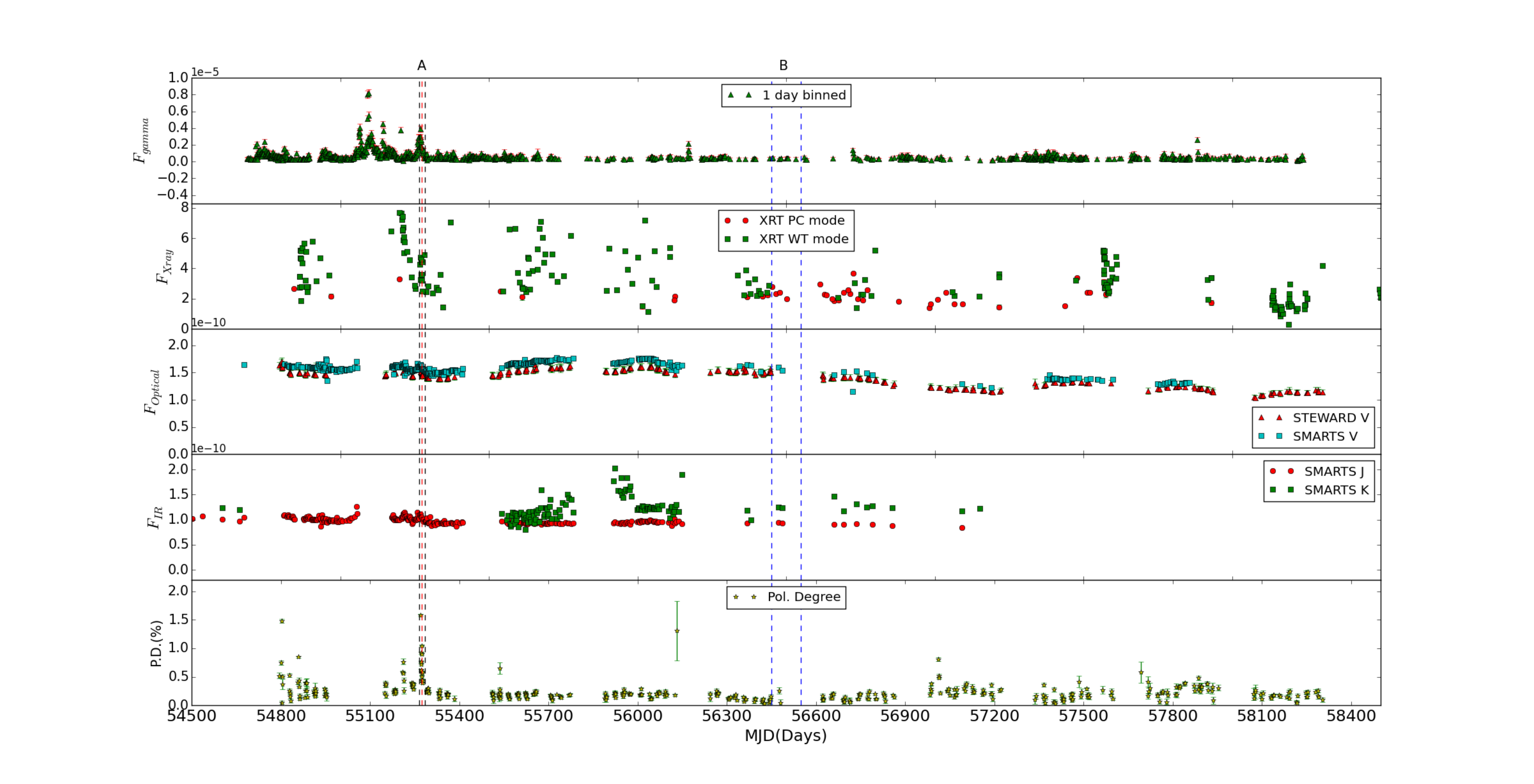}
\caption{Multi-wavelength light curves of the source 3C 273. The panels and the vertical
lines are as in Fig.\ref{figure-1}}
\label{figure-4}
\end{figure*}

\begin{figure*}
\hbox{
\includegraphics[scale=0.45]{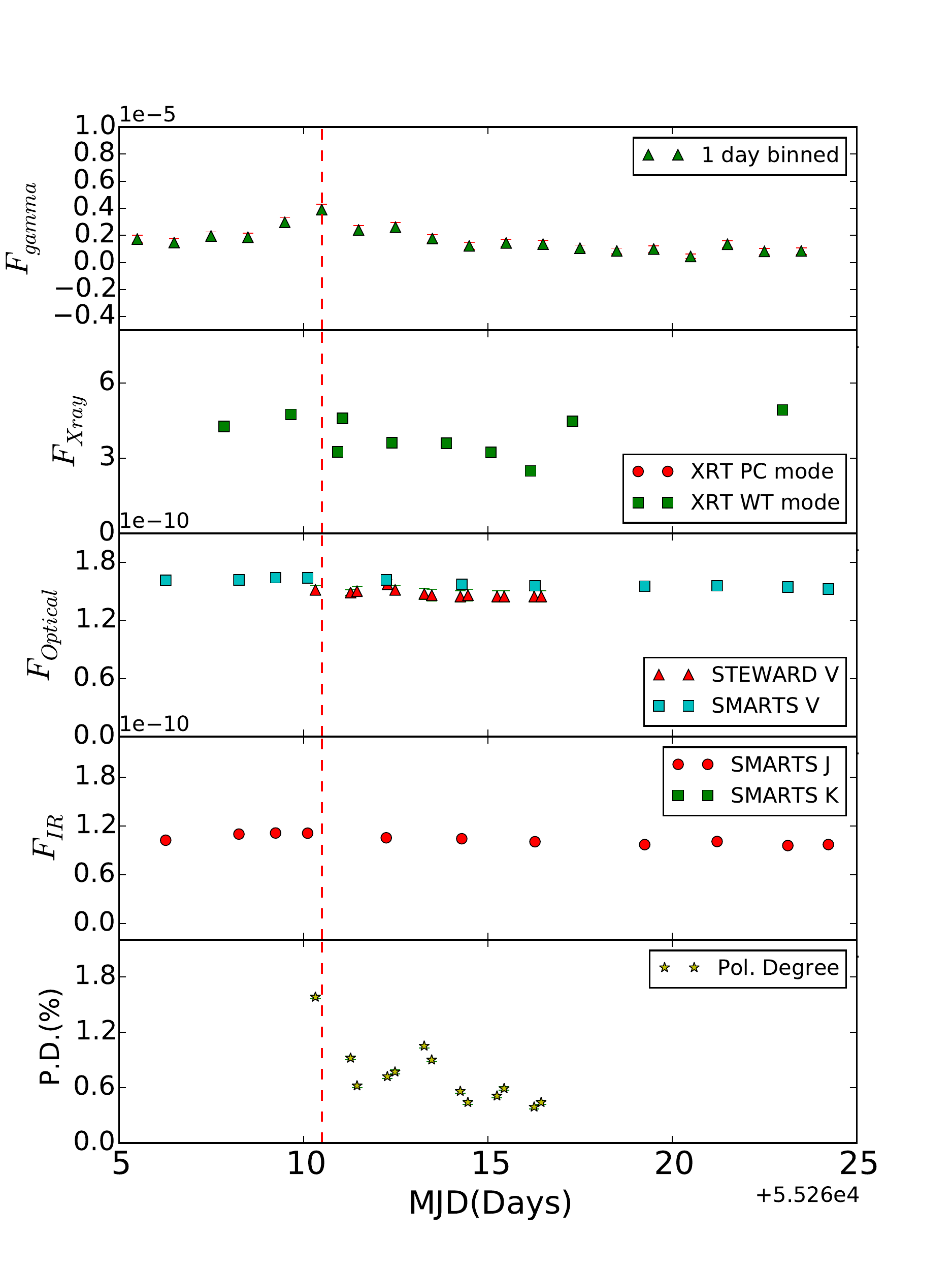}
\includegraphics[scale=0.45]{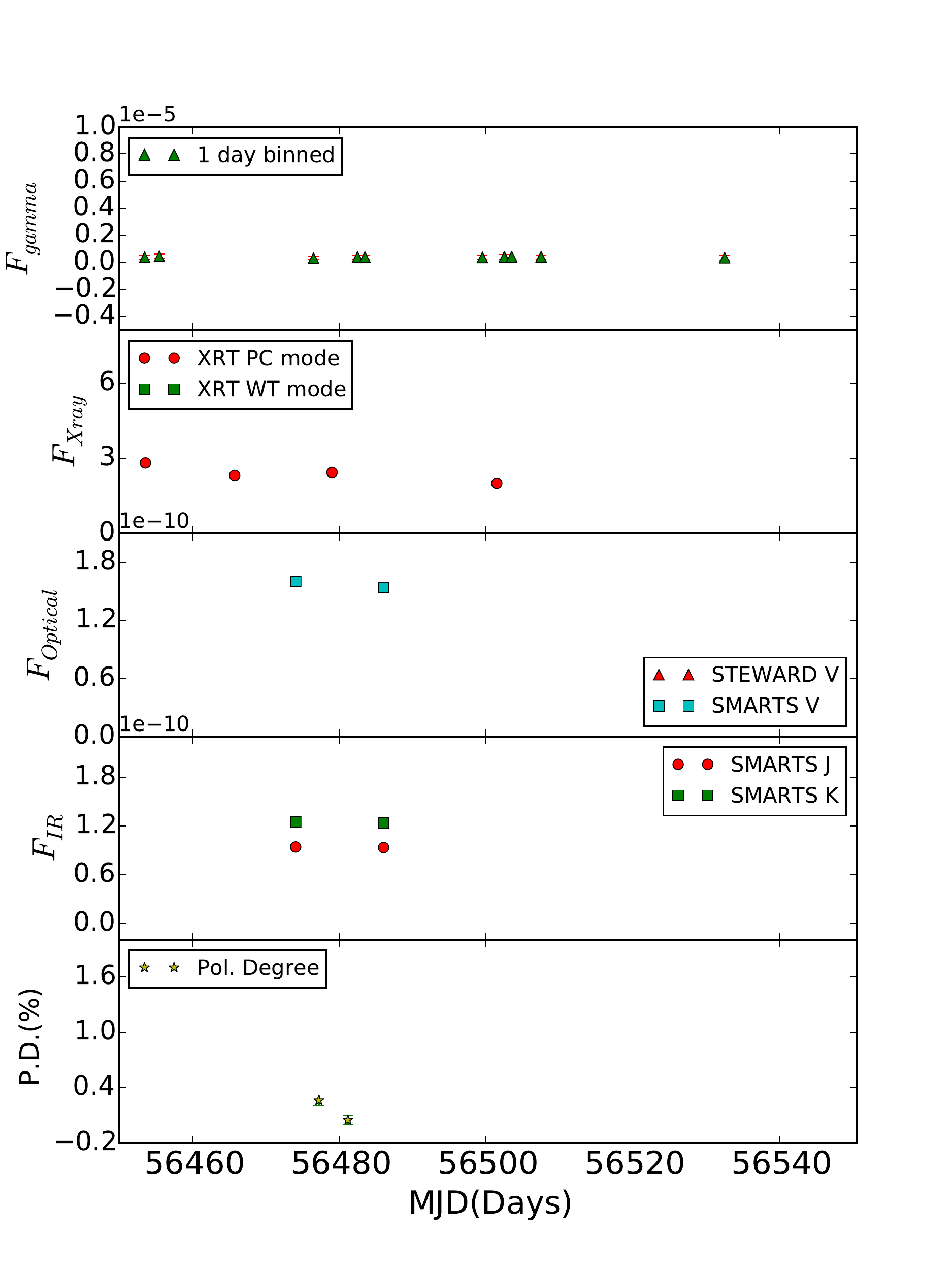}
}
\caption{Light curves of 3C 273 for epoch A (left panel) and epoch B (right panel).
The $\gamma$-ray fluxes are in units of 10$^{-5}$ ph cm$^{-2}$ s$^{-1}$ and the 
optical and IR fluxes are in units of 10$^{-10}$ erg cm$^{-2}$ s$^{-1}$. The 
vertical dotted line shows the peak of the $\gamma$-ray flare.}
\label{figure-5}
\end{figure*}

\begin{figure*}
\hspace*{-3cm}
\includegraphics[width=1.3\textwidth]{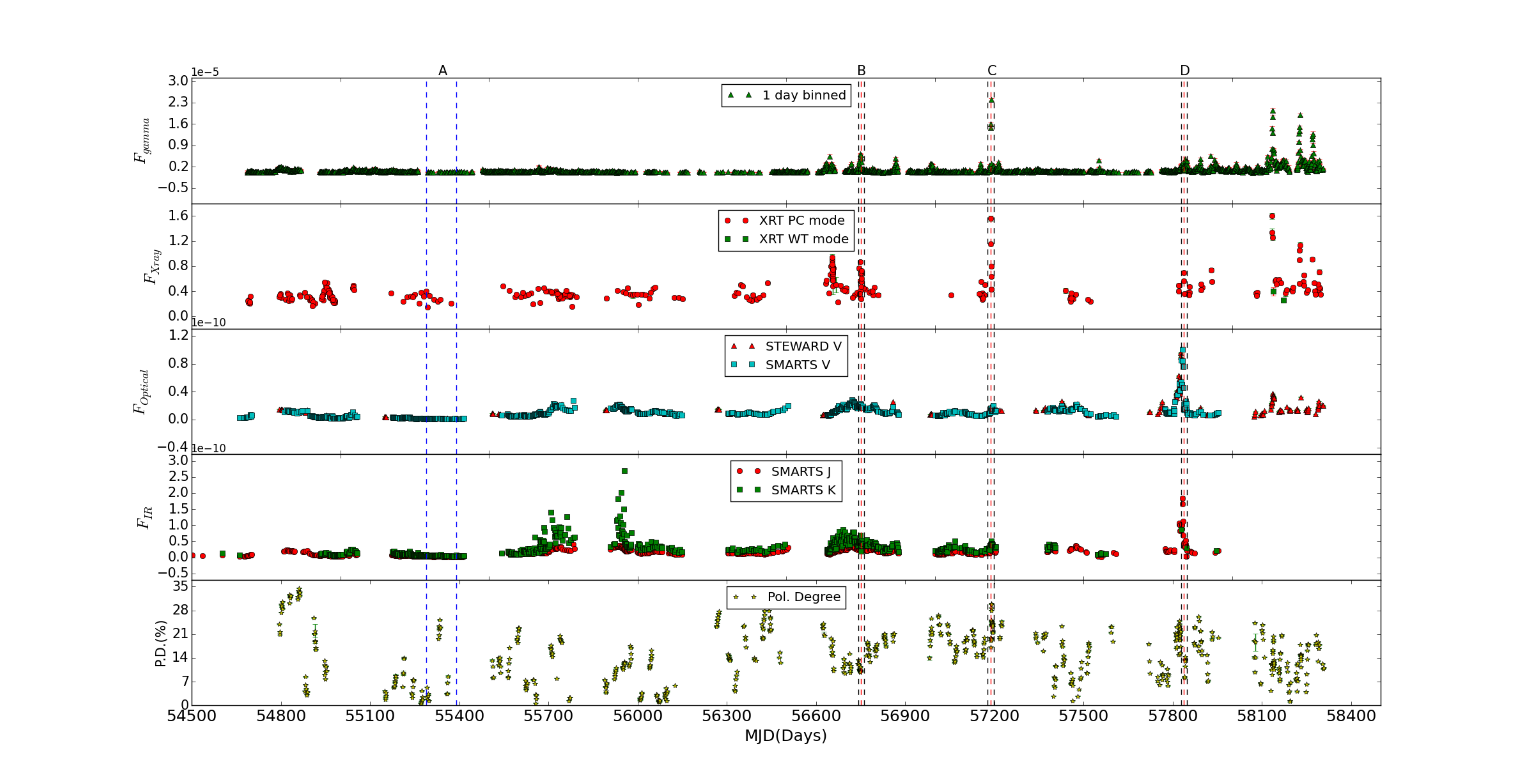}
\caption{The light curves of the source 3C 279 in different wavelengths. The panels and
the dashed lines are as in Fig. \ref{figure-1}.}
\label{figure-6}
\end{figure*}  

\begin{figure*}
\vbox{
\hbox{
\includegraphics[scale=0.45]{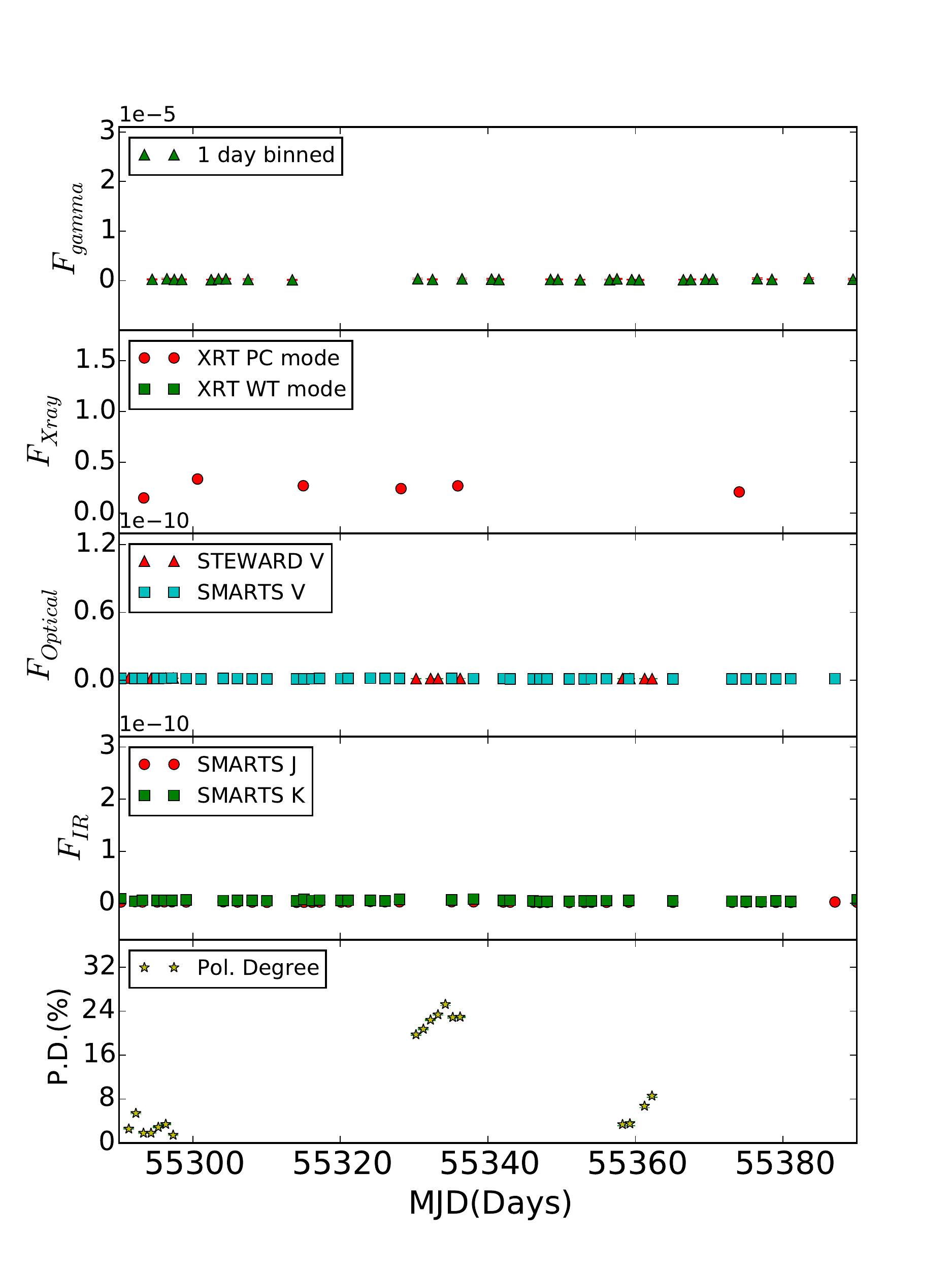}
\includegraphics[scale=0.45]{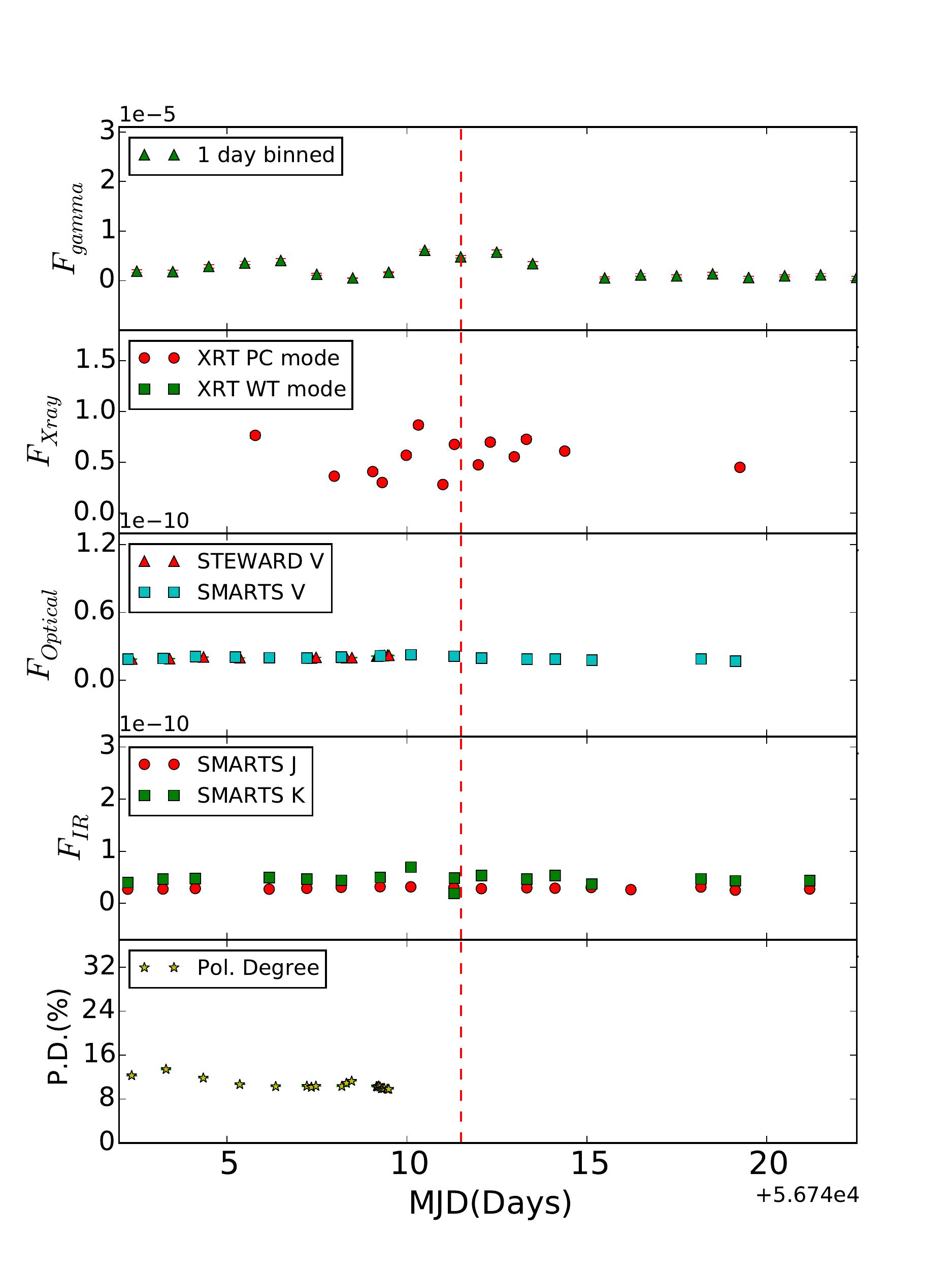}
}
\hbox{
\includegraphics[scale=0.45]{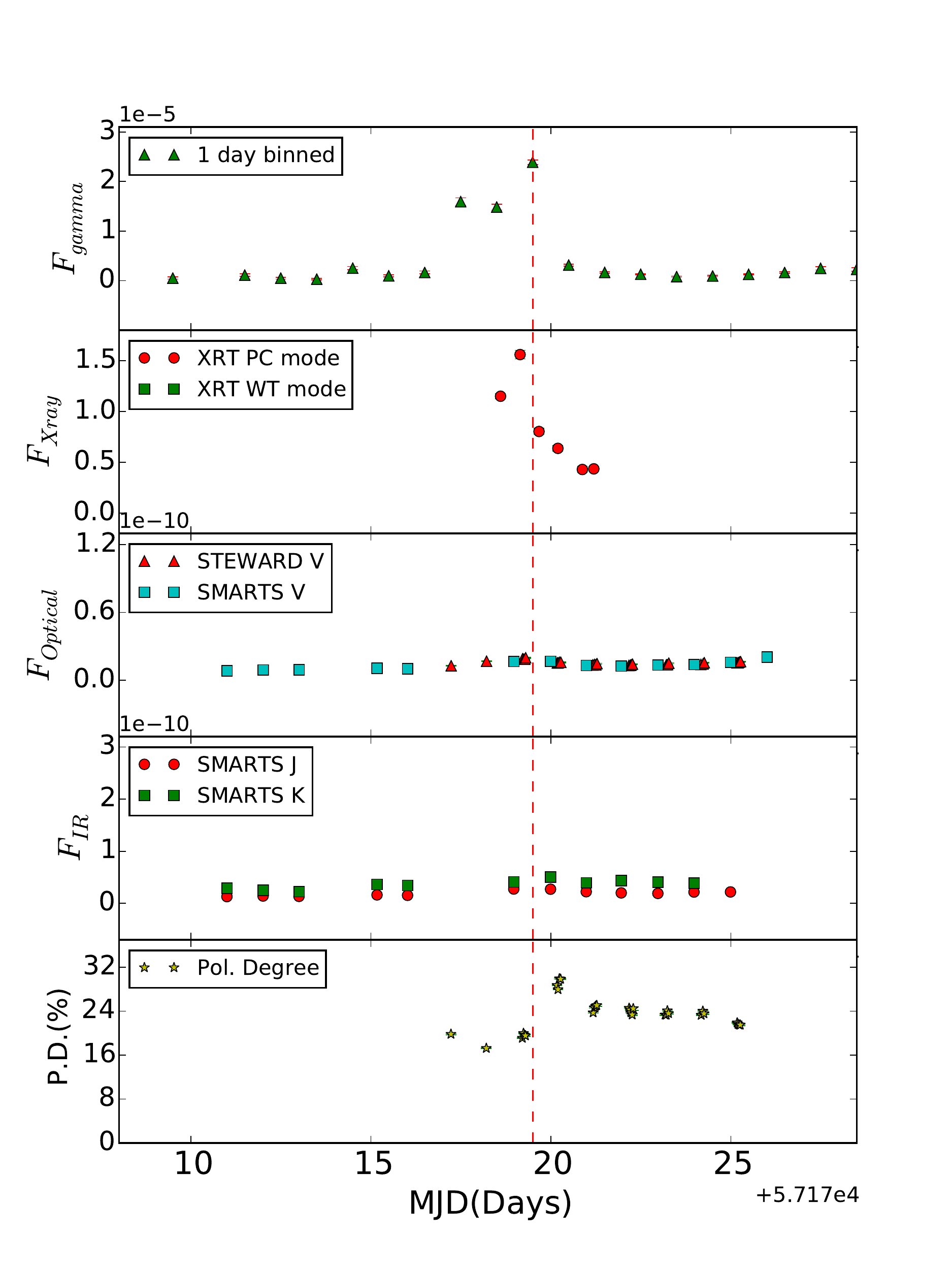}
\includegraphics[scale=0.45]{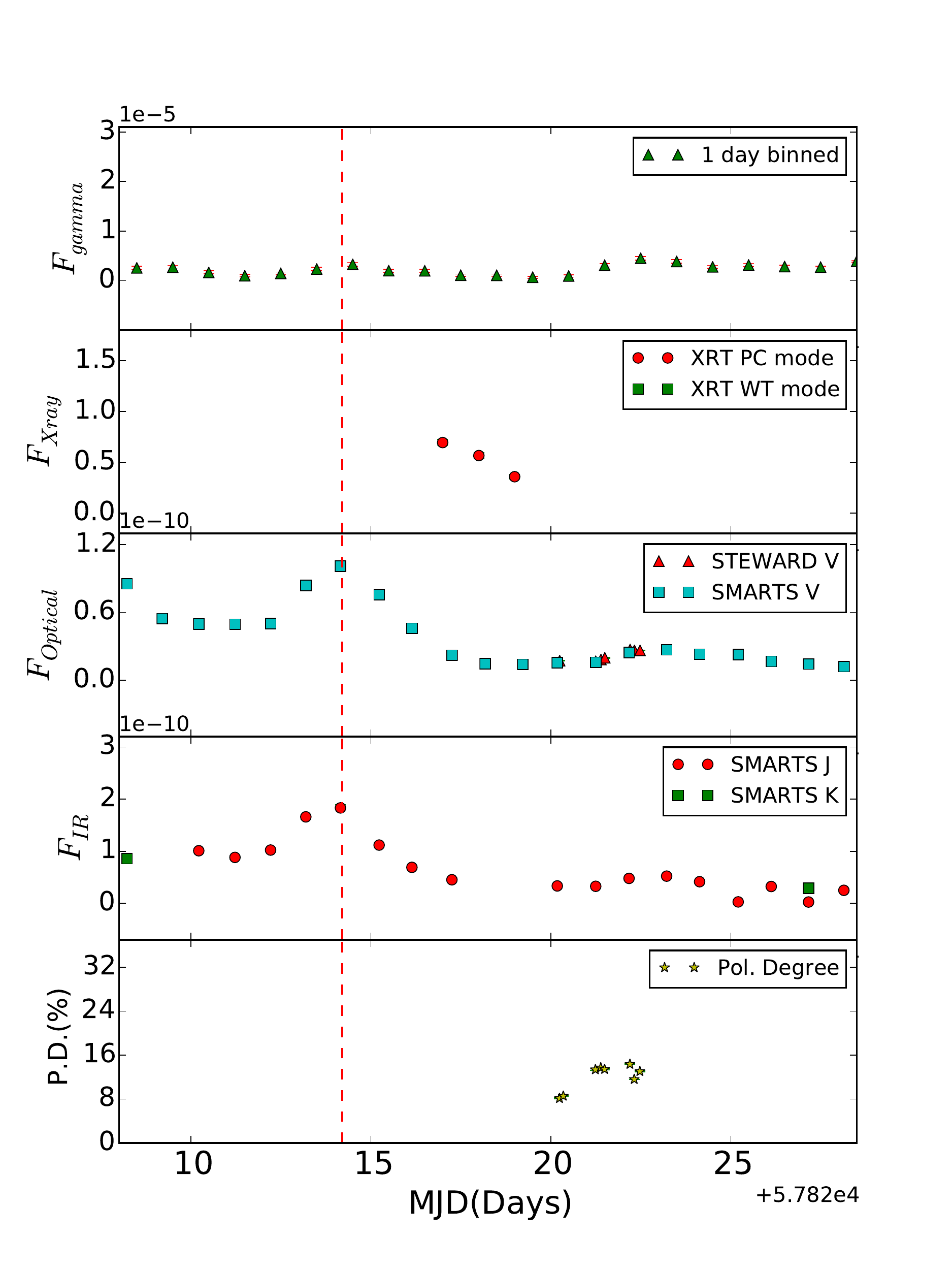}
}
}
\caption{Multi-wavelength light curves of the source 3C 279 for epoch A (top left), 
epoch B (top right), epoch C (bottom left) and epoch D (bottom right). The optical 
and IR light curves have units of 10$^{-10}$ erg cm$^{-2}$ s$^{-1}$, while 
the $\gamma$-ray light curves have units of 10$^{-5}$ ph cm$^{-2}$ s$^{-1}$. The 
dashed lines indicate the peak of the optical/$\gamma$-ray
flare.}
\label{figure-7}
\end{figure*}

\begin{figure*}
\hspace*{-3cm}
\includegraphics[width=1.3\textwidth]{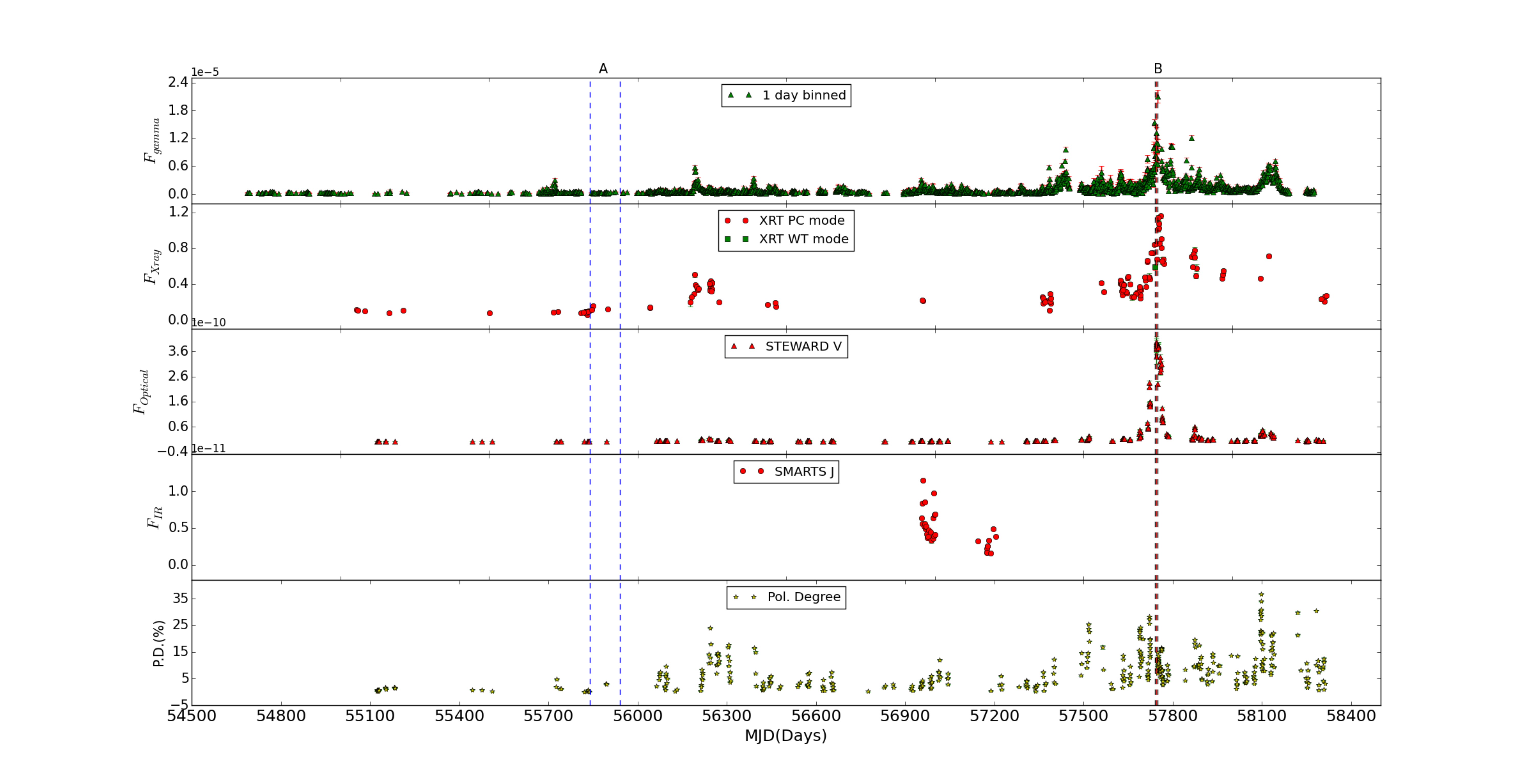}
\caption{Long term light curves of the source CTA 102 in different wavelengths. Details in this figure are similar to that of Fig. \ref{figure-1}.}
\label{figure-8}
\end{figure*} 

\begin{figure*}
\hbox{
\includegraphics[scale=0.45]{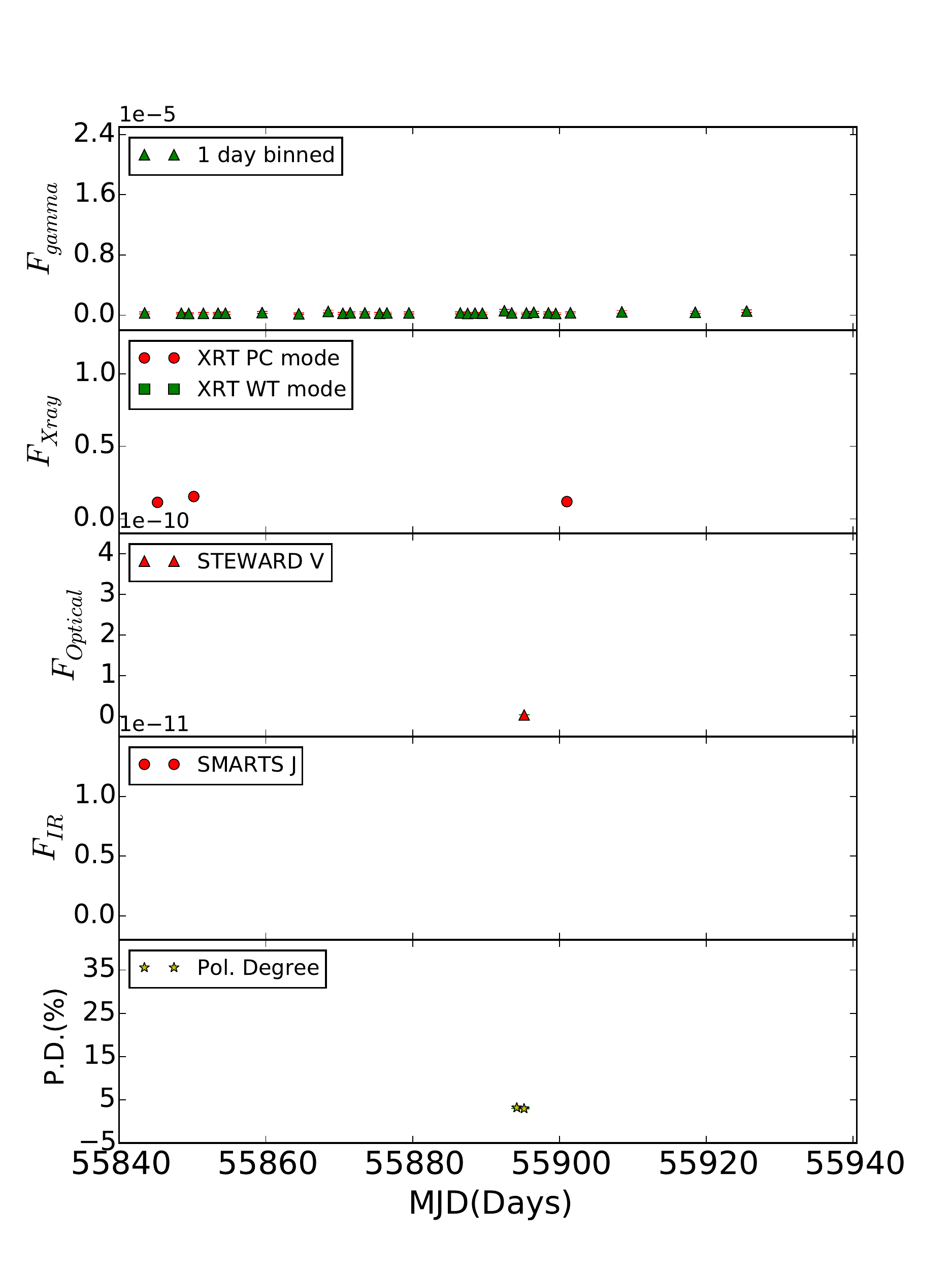}
\includegraphics[scale=0.45]{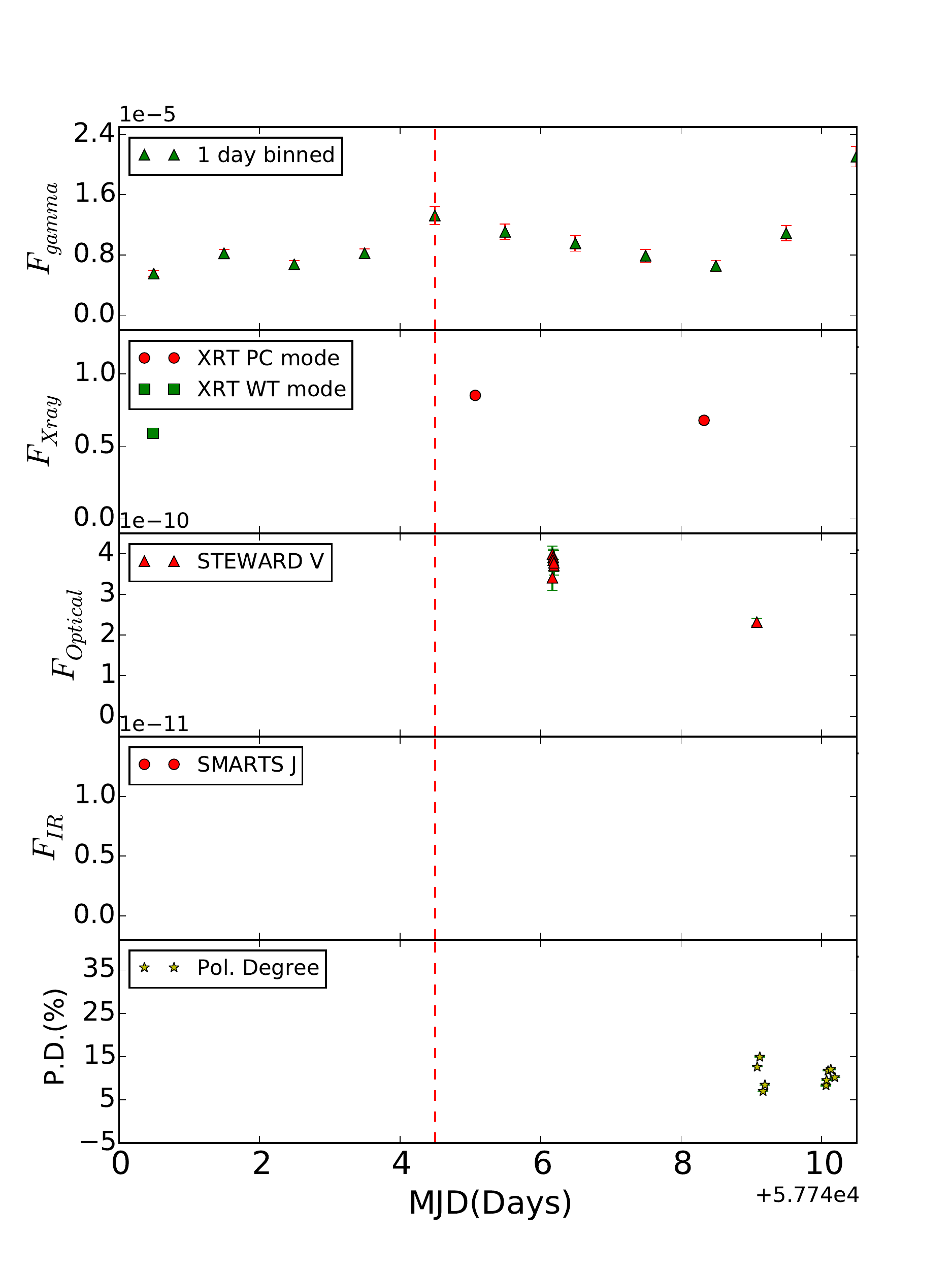}
}
\caption{The left and right panels show the multi-wavelength light curves of the source CTA 102 for
epoch A and B respectively. The dashed line shows the peak of the $\gamma$-ray flare. The optical and
$\gamma$-ray fluxes are in units of 10$^{-10}$ erg cm$^{-2}$ s$^{-1}$  and 10$^{-5}$ ph cm$^{-2}$ s$^{-1}$ respectively.}
\label{figure-9}
\end{figure*}

\section{Reduction of Multi-wavelength data} \label{sec:data}
Our aim in this work is to characterize the connection between optical and $\gamma$-ray
flux variations in FSRQs. Therefore, primarily data in both the optical and $\gamma$-ray bands
are needed. However, for broad band SED modelling, data from other wavelength regions are
also required. Thus, for this work we used all the publicly available data in the IR, optical,
UV, X-rays and $\gamma$-rays that span a period of 10 years between 08 August 2008 and
08 August 2018. Optical polarimetric data if available during the above period was also used.

\subsection{UV-Optical and IR data}
We used the {\it Swift}-UV-Optical Telescope (UVOT) for data in the UV and optical bands. 
They were analyzed using the online tool\footnote{https://www.ssdc.asi.it/cgi-bin/swiftuvarchint}. To generate the light curve the magnitudes thus obtained and uncorrected for Galactic reddening were then converted to fluxes 
using the zero points taken from \cite{2011AIPC.1358..373B}. However corrections due to galactic absorption were applied to generate the average data points for SED analysis. In addition to the 
optical data from {\it Swift}-UVOT, we also used optical data in the V-band from
both SMARTS and the Steward 
Observatory, while the IR observations in the 
J and K-bands were taken from SMARTS. Optical polarization data whereever
available were taken from the Steward Observatory. The details of the instrument and
the data reduction procedures for SMARTS can be found in \cite{2012ApJ...756...13B}, while 
the details on the observations and reductions of data from Steward Observatory can be found in 
\cite{2009arXiv0912.3621S}.

\subsection{X-ray Data}
We used data from the X-ray telescope (XRT) on board {\it Swift} covering the 
energy range from 0.3 $-$ 10 keV \citep{2005SSRv..120..165B,2004ApJ...611.1005G}.
The data that spans about 10 years and covering the period August 2008 to August 2018 were taken from the archives at HEASARC \footnote{https://heasarc.gsfc.nasa.gov/docs/archive.html}. 
The data were analyzed by the instrument pipleine following standard procedures. For light curve analysis, we used the data 
collected using both  window timing (WT) and photon counting (PC) modes. For spectral analysis of the sources PKS 1510$-$089, 3C 279 and CTA 102 we used the data collected only from 
the PC mode, while, for the source 3C 273 we used PC mode data for the quiescent state 
but for $\gamma$-ray flaring state we used the WT mode data due to the non availability 
of PC mode data. The  XRT data were processed with the {\tt xrtpipeline} task using the 
latest CALDB files available with version HEASOFT-6.24. We used the standard grade selection 0-12 and
the calibrated and cleaned events were added to generate the energy spectra. For PC mode, we  
extracted the source spectra from a circular region of radii 60$^{\prime\prime}$, 
and  the background spectra were selected from the region of radii 80$^{\prime\prime}$ away from the source. In WT mode, for the source  we used a circular region of 60$^{\prime\prime}$ radii 
while for the background we used the region between 80$^{\prime\prime}$ and 
120$^{\prime\prime}$ radii centered around the source. We combined the exposure maps using {\tt XIMAGE}  
and created  the ancillary response files using {\tt xrtmkarf}. For fitting the 
data within XSPEC \citep{1996ASPC..101...17A}, we used an absorbed simple power 
law model with the Galactic neutral hydrogen column densities of 
$N_{\rm H}$ = 6.89$\times$ $10^{20}$ $cm^{-2}$, 2.21$\times$ $10^{20}$ $cm^{-2}$, 
4.81$\times$ $10^{20}$ $cm^{-2}$ and 1.68$\times$ $10^{20}$  $cm^{-2}$ from \cite{2005A&A...440..775K} 
for the sources PKS 1510$-$089, 3C 279, CTA 102 and 3C 273 respectively. We used $\chi^2$ statistics
within XSPEC and the uncertainties were calculated at the 90\% confidence level.

\subsection{$\gamma$-ray data}
We analyzed 10 years of $\gamma$-ray data from the LAT instrument on board {\it Fermi} 
during the period 08 August 2008 to 08 August 2018 to generate the one day binned 
$\gamma$-ray light curve. The LAT is an imaging high energy $\gamma$-ray 
telescope, sensitive in the energy range from 20 MeV$-$300 GeV. The 
field of view of the LAT instrument is 20\% of the sky and it scans continuously, 
covering the whole sky every three hours \citep{2009ApJ...697.1071A}. We used 
{\it fermipy} to analyze the 10 years of $\gamma$-ray data. {\it Fermipy} is a 
python software package that provides a high-level interface for LAT data 
analysis \citep{2017ICRC...35..824W}. We used Pass 8 data for the analysis 
where the photon-like events are classified as 'evclass=128, evtype=3' with energy 
range 0.1$\leqslant$E$\leqslant$300 GeV. A circular region of interest (ROI) of 
$15^{\circ}$ was chosen with a zenith angle cut of $90^{\circ}$ in order to 
remove Earth limb contamination. We used the latest isotropic model "iso$\_$P8R2$\_$SOURCE$\_$V6$\_$v06" 
and the Galactic diffuse emission model "gll$\_$iem$\_$v06" for the analysis. The recommended criteria 
"(DATA$\_$QUAL>0)\&\&(LAT$\_$CONFIG==1)" was used for the good time interval selection. In 
the generation of $\gamma$-ray light curves, we considered the source to be detected
if the test statistics (TS) is $>$ 9 which corresponds to a 3$\sigma$ detection \citep{1996ApJ...461..396M}.\\

\section{Analysis}
\subsection{Multi-wavelength light curves}
Analysis for the presence or absence of correlation between optical and 
$\gamma$-ray flares requires identification of flares in optical and/or 
$\gamma$-ray light curves. Due to large gaps and/or less number of 
points in the optical light curves it is not possible to automatically
identify epochs (through cross-correlation analysis) on the presence
or absence of correlated optical and $\gamma$-ray flux variations. 
Therefore, flares for detailed analysis were selected visually as follows.
Multi-wavelength light curves that span the 10 year period were first generated for each object. In that, optical and $\gamma$-ray light curves were visually inspected for the presence of sharp peaks
above their base flux levels. Once identified, expanded multiwavelength light curves were generated for a total duration of 20 days, centered at the optical and/or $\gamma$-ray flares. In an epoch when a $\gamma$-ray flare or an optical flare is identified, we imposed the condition of 
having data in multiple wavelengths such as $\gamma$-rays, X-rays, UV, optical 
and IR so as to constrain both the low energy and high energy hump in the
SED analysis. These conditions lead to the identification of few flares. Of 
these we concentrated only on some epochs for each object.
\subsubsection{PKS 1510$-$089}
The multi-wavelength light curves that include $\gamma$-ray, X-ray, UV, optical and IR are given in 
Fig. \ref{figure-1}. The figure also includes polarization measurements. Inspection of 
the light curves indicates that the source has displayed varied activity levels that
includes both flaring and quiescent periods. From visual inspection of the 
light curves we selected 6 epochs (A, B, C, D, E and F) in the source for studying the correlations between optical and $\gamma$-ray
variations. These epochs were identified by the presence of optical and/or $\gamma$-ray flares in the light curves and a quiescent
state in both the optical and $\gamma$-ray bands. A summary of these epochs is given in Table \ref{table-2} and the multi-wavelength
light curves covering a shorter duration during these epochs are shown in 
Fig. \ref{figure-2} and Fig. \ref{figure-3}. More details on these
six epochs are given below:

\noindent {\bf Epoch A:} During this epoch, the $\gamma$-ray has increased in flux
by a factor of about 10, while the optical and the IR J and K-band fluxes have not shown any
variability and are indeed steady. There is also a hint that the X-ray flux from 
the source is non-variable, however, due to the lack of data during part of the 
$\gamma$-ray flare, no conclusive statement could be made on the nature of
X-ray flux variations. The optical polarization too has not
shown noticeable variability during the steady optical/IR brightness
state of the source. We conclude that in this epoch we observed a $\gamma$-ray flare
with no optical counterpart.

\noindent {\bf Epoch B:} During this epoch, the optical flux has increased 
by a factor of 6, while the flux variations in the IR band are at a reduced level.
There is also a hint of a very low amplitude $\gamma$-ray flare during the
peak of the optical flare, but it is very small. The lack of X-ray data
and optical polarization data during the epoch of the optical flare
prevents us to make any statement on the nature of X-ray variations as
well as the degree of optical polarization during this epoch. Thus in this
epoch the source has shown correlated optical and $\gamma$-ray flux variations,
though the amplitude of variations in the $\gamma$-ray band is much lower than
that of the optical and IR bands.

\noindent {\bf Epoch C:} The flux variations noticed in this epoch is similar to
that observed during epoch A. A minor flare is observed in the $\gamma$-ray band, 
but the source is stable in the X-ray, optical and
IR bands. We do not have optical polarization data during the $\gamma$-ray
flare for an idea on the nature of optical V-band polarization. Thus, during
this epoch, the source has shown a $\gamma$-ray flare without a counterpart in the low energy
X-ray, optical and IR bands.

\noindent {\bf Epoch D:} During this period, the source is in the quiescent state
in all the energy bands analyzed here. 

\noindent {\bf Epoch E:} During this epoch, the source has shown a strong 
$\gamma$-ray flare, however, such a flare is not seen
in the X-ray, optical and IR bands. Here too, optical polarization data is not available
during the period of the $\gamma$-ray flare. Thus, in this epoch, the source
has shown a $\gamma$-ray flare without similar flaring in the other wavelengths such as
X-rays, optical and IR. 

\noindent {\bf Epoch F:} A weak $\gamma$-ray flare is seen in the source
during this period. Simultaneous to the $\gamma$-ray flare, there is an 
indication of a minor optical flare which is also
accompanied by an increase in the degree of optical polarization.  
There is paucity of X-ray
and IR data during the peak of the $\gamma$-ray flare. Thus, the optical and $\gamma$-ray 
flux variations are closely correlated during this epoch.

\subsubsection{3C 273}
We show in Fig. \ref{figure-3} the multi-wavelength light curves for the source 
3C 273. The source is mostly quiescent during the period 2008 to 2018 August 
except for few instances where it has flared in the $\gamma$-ray band. We 
identified two epochs in this source for studying the correlation between 
optical and $\gamma$-rays. The details of these two epochs are given in 
Table \ref{table-2}. They are also marked in Fig. \ref{figure-4}, and an expanded 
version of these two epochs is given in Fig. \ref{figure-5}.

\noindent{\bf Epoch A:}
There is a prominent $\gamma$-ray flare during this epoch,  wherein the $\gamma$-ray flux has
increased by a factor of two at the peak of the $\gamma$-ray flare. During the peak of 
the $\gamma$-ray flare, the X-ray,  optical and IR brightness do not show significant changes. 
Also, the source lacks optical polarization data during the peak of the $\gamma$-ray flare. The source has thus shown a $\gamma$-ray flare without an optical counterpart in this epoch.

\noindent{\bf Epoch B:}
During this epoch the source is found to be in the quiescent state in all
the wavebands namely $\gamma$-rays, X-ray, optical and IR. The source is also
weakly polarized in the optical V-band during this period.

\subsubsection{3C 279}
We show in Fig. \ref{figure-6} the multi-wavelength light curves. From visual inspection, we 
identified four epochs in this source for studying the correlation between 
optical and $\gamma$-ray flux variations. We give in Table \ref{table-2} the summary of those 
four epochs. An expanded view of the multi-wavelength flux variations in the source
is shown in Fig. \ref{figure-7}. Below, we summarize the salient aspects of these four epochs.

\noindent{\bf Epoch A:} In this epoch the source is in the quiescent state. In the one day
binned $\gamma$-ray light curve the source is below the detection limit for
many days during this 100 days period. Also, in the X-ray, optical and IR bands, 
the source is non-variable during this period. However, during
the middle of this epoch, the optical polarization increased by a
factor of about 5 from $\sim$6\% to $\sim$25.8\%. During this period of increased
optical polarization, the source did not show flux variations in any of the
bands.

\noindent{\bf Epoch B:} During this epoch, the source has shown a minor flare in the $\gamma$-ray band
with no corresponding flare in the optical, IR and X-ray bands. Polarization
data is not available during the peak of the $\gamma$-ray flare thereby making it
impossible to know the polarized nature of the source. By comparing the multi-wavelength light curves during this epoch, we
conclude that the source showed a $\gamma$-ray flare without an optical counterpart.  

\noindent{\bf Epoch C:} During this epoch, a strong $\gamma$-ray flare was observed wherein
the $\gamma$-ray flux increased by a factor of about 3. During the peak
of the $\gamma$-ray flare, X-ray too showed a flare, however, in the
optical and IR bands, the source was found to be stable with no signs
of flux variability. An interesting behaviour displayed by this source is an
apparent negative correlation of $\gamma$-ray and  X-ray flux variations to
the optical polarization. During the epoch when the $\gamma$-ray and
X-ray were at their peaks, the optical polarization was low, and it
gradually increased when the X-ray and $\gamma$-ray fluxes declined.

\noindent{\bf Epoch D:} During this epoch, the source showed a prominent optical 
and IR flare. The flare was found be be asymmetric with a fast rise and slow decay. During the
epoch of the optical and IR flare the source did not show any variation in 
the $\gamma$-ray band.  Due to the lack of polarization data during the peak of 
the optical and IR flare, we could not make any statement on the optical polarization state during
the time of the optical and IR flare. Thus this epoch is a clear example of the source
showing an optical flare without a $\gamma$-ray counterpart. 

\subsubsection{CTA 102}
The source was found to be in a steady and low brightness state during most of 
the time between 2008 August to 2018 August, except for a spectacular
$\gamma$-ray flare in the beginning of 2016. The multi-wavelength light curves are shown in Fig. \ref{figure-8}. We have 
identified 2 epochs in this source for studying the correlation between optical
and $\gamma$-ray variations. A summary of these two epochs is given in Table \ref{table-2} and expanded 
plots of these two epochs are shown in Fig. \ref{figure-9}. More details on these two epochs are given below.

\noindent{\bf Epoch A:}
During this epoch the source was in the quiescent state in all the wavebands
considered in this work.

\noindent{\bf Epoch B:}
The source showed a major $\gamma$-ray flare during this epoch. This
flaring in the $\gamma$-ray band was also accompanied by flaring behaviour in the  
X-ray and optical wavelengths. The nature of IR flux during this period is uncertain 
due to the non-availability of IR data during this flaring period. Thus, during
this epoch, the source showed correlated flux variations in the optical and 
$\gamma$-ray bands.

\subsection{Spectral Variations}
\begin{figure*}
\includegraphics[scale=1.00]{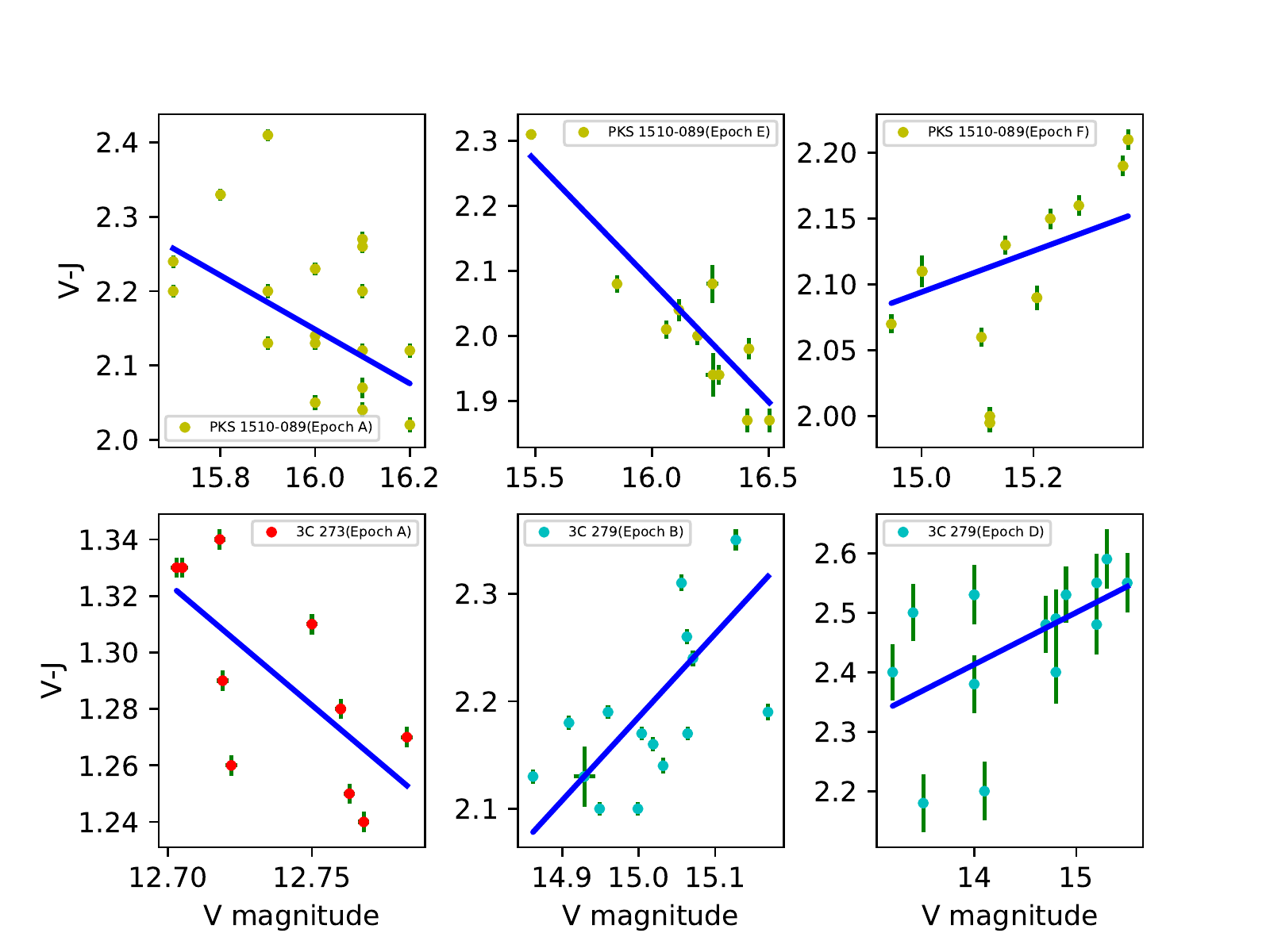}
\caption{Colour-magnitude relations. From the left, the top panels are for the 
source PKS 1510$-$089, epochs A, E and F.  In the bottom the left panel is for the 
epoch A of 3C 273, while the other two panels are for the epochs B and D of 3C 279.}
\label{figure-10}
\end{figure*}

Blazars show spectral variations in addition to flux variations. To characterize the 
spectral variability of the blazars studied here, we looked for variations in the V$-$J 
color against the V-band brightness. This spectral analysis was  done for all the 
epochs in the sources PKS 1510$-$089, 3C 273 and 3C 279 for which correlations 
between optical and $\gamma$-ray flux variations 
were explored in Section 4.1. 
Such spectral analysis was not carried out for the source CTA 102 due to the 
lack of J-band data. Spectral variations were characterized by linear least squares
fit to the colour-magnitude diagram by taking into account the errors in both the colour
and magnitude. A source is considered to have shown colour variation if 
the Spearman rank correlation coefficient is $> 0.5$ as well as $<$ $-$0.5 
with the null hypothesis probability P $<$ 0.05. In the source PKS 1510$-$089, we found a "redder when brighter" (RWB) 
trend at epochs A and E, however, during epoch F we observed a "bluer 
when brighter" (BWB) behaviour. In the source 3C 273, during epoch A, we found the source
to show a RWB trend. For the source 3C 279, we found BWB behaviour during epochs 
B and D. This indicates that the spectral
variations shown by FSRQs are complex and a FSRQ may not show the same 
spectral variability pattern at all times. The colour magnitude diagram for the sources PKS 1510$-$089, 3C 273 and 3C 279 are shown in Fig. \ref{figure-10}. 

\subsection{$\gamma$-ray spectra}
To study the intrinsic distribution of electrons in the jets that are involved 
in the $\gamma$-ray emission process, we generated $\gamma$-ray spectra for all 
the selected epochs (as detailed in Section 4.1) of the four sources. We fitted 
the $\gamma$-ray spectra with the two models namely (i) the  simple power law (PL) 
model and (ii) the log-parabola (LP) model. For PL model we used the following
\begin{equation}
dN(E)/dE=N_{\circ}(E/E_{\circ})^{\Gamma_{PL}}
\end{equation}
where $N_{\circ}$ is the normalization of the energy spectrum and 
$E_{\circ}$ is the scaling factor and $\Gamma_{PL}$ is the photon index.\\
The LP model has the following form \citep{2012ApJS..199...31N}  
\begin{equation}
dN(E)/dE=N_{\circ}(E/E_{\circ})^{-\alpha-\beta ln(E/E_{\circ})}
\end{equation}
where, dN/dE is the number of photons in cm$^{-2}$ s$^{-1}$ MeV$^{-1}$,  $\alpha$ 
is photon index at $E_{\circ}$, $\beta$ is the curvature index that defines the 
curvature around the peak, E is the energy of the $\gamma$-ray photon, N$_{\circ}$ 
is the normalization and E$_{\circ}$ is the 
scaling factor. 

\begin{figure*}
\hbox{
\includegraphics[scale=0.5]{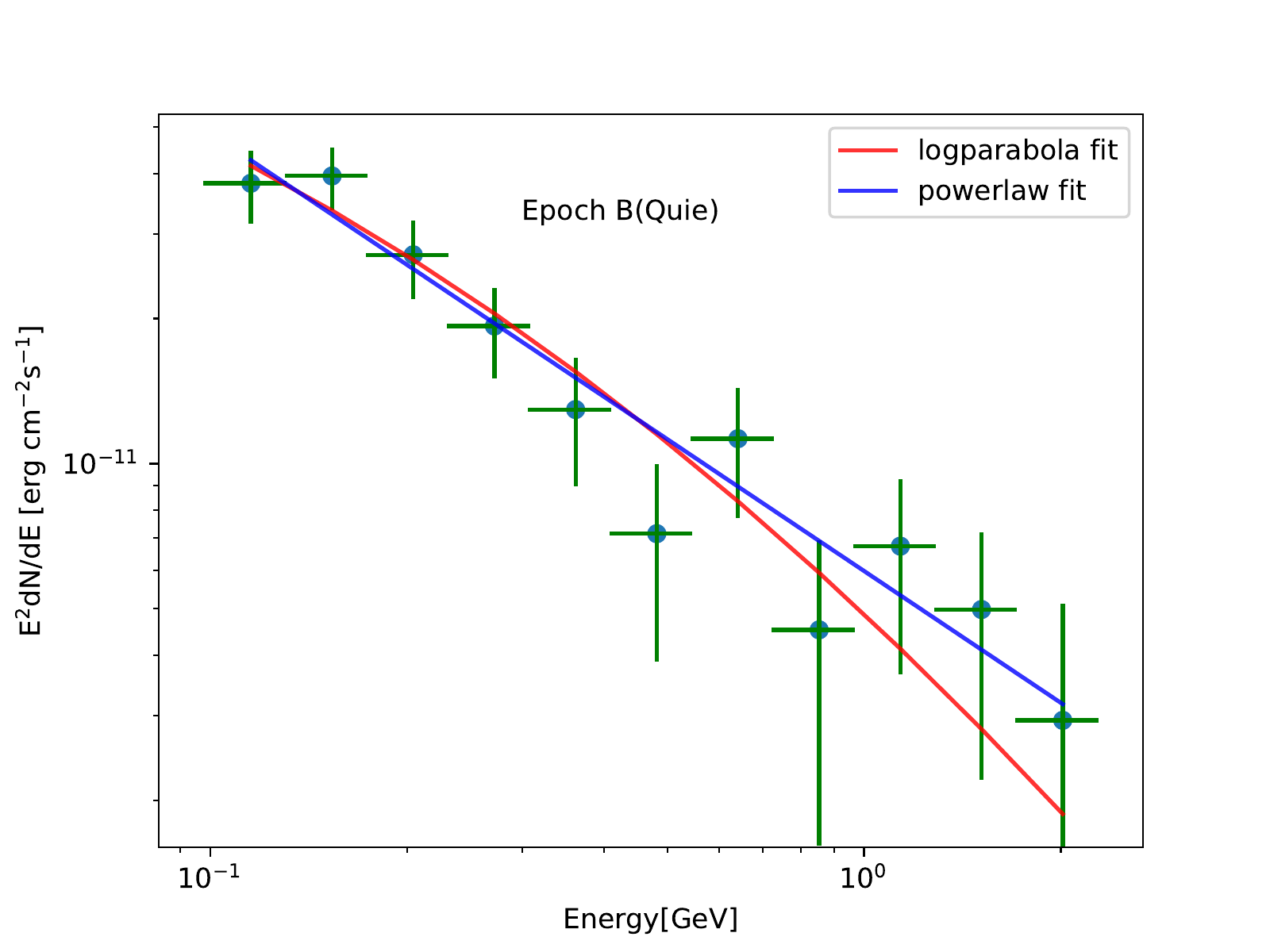}
\includegraphics[scale=0.5]{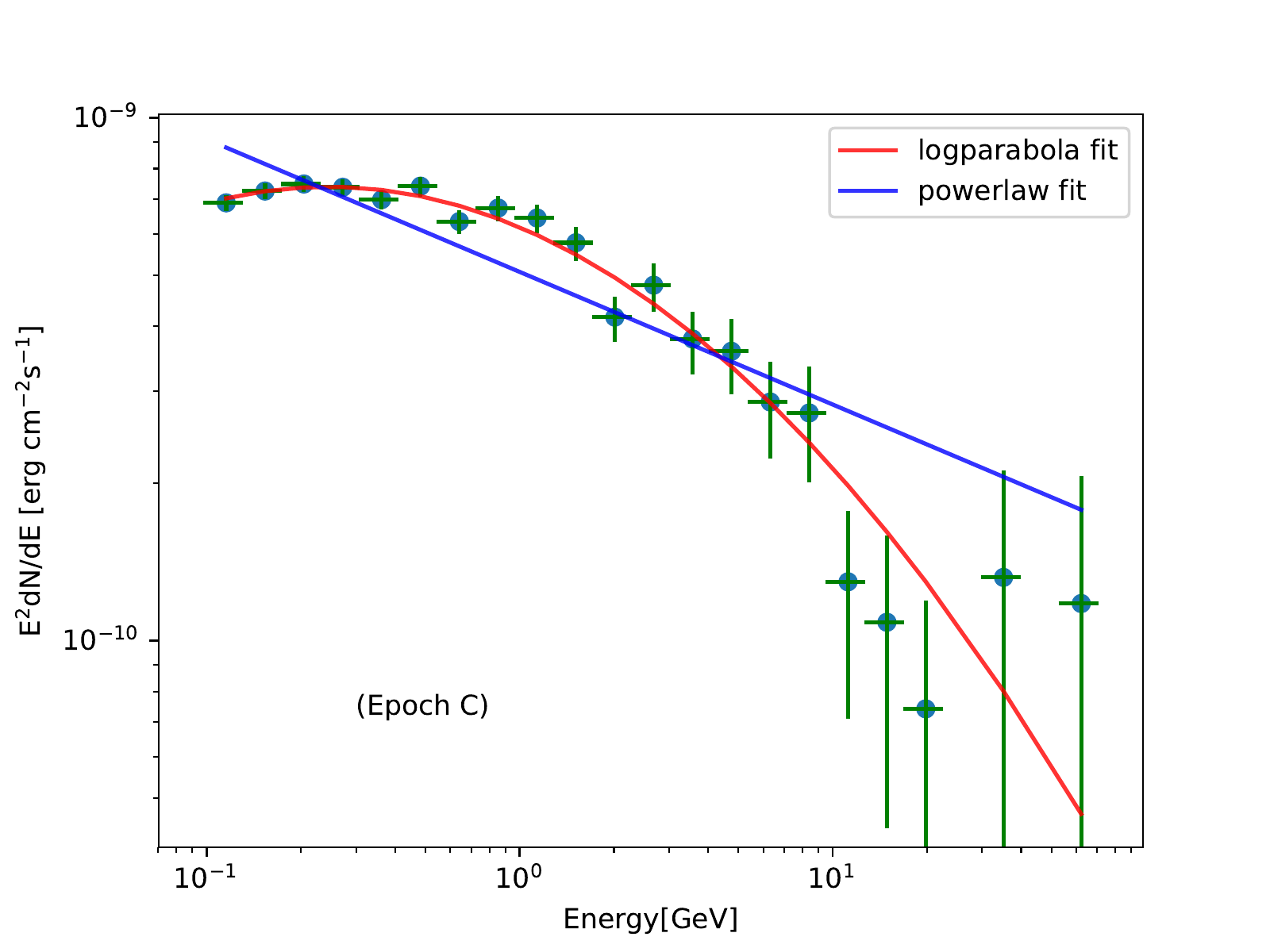}
    }
\caption{Observed and model fits to the $\gamma$-ray spectra of the source 3C 273 
for epoch B (left panel) and of 3C 279 for epoch C (right panel).}
\label{figure-11}
\end{figure*}

To test the model that well describes the $\gamma$-ray spectra (PL against LP), 
as well as the presence of curvature, we used the maximum likelihood estimator {\tt gtlike}. Following \cite{2012ApJS..199...31N}, we calculated the curvature of the test statistics as $TS_{curve}$ = 2(log $L_{LP}$ - log $L_{PL}$). Here L represents the likelihood function.
We used the threshold $TS_{curve}$ > 16 for the presence of 
a statistically significant curvature in the $\gamma$-ray spectra, (at the 4$\sigma$ level; \citealt{1996ApJ...461..396M}). We found that for all the epochs, the $\gamma$-ray spectra is well fit by the LP model except for the quiescent epoch of the source 3C 273 which is well fit by the PL model. The results of the $\gamma$-ray spectral analysis are given in 
Table \ref{table-3}. One example of the PL model that best fits the data (for the source 3C 273 at Epoch B) and
another example of a LP model that best fits the data (for the source 3C 279 at Epoch C) is shown in Fig. \ref{figure-11}. 

\begin{table*}
\caption{Details of the PL (Eq. 1) and LP (Eq. 2) model fits for the different epochs of the 
sources PKS 1510$-$089, 3C 273, 3C 279 and CTA 102. $\Gamma_{PL}$ is the photon index from PL fitting, $\alpha$ and $\beta$ the photon index and curvature index from LP fit to the spectra, TS is the test statistics, Log L is the log-likelihood, and $TS_{curve}$ is the curvature of the test statistics defined as 2(log $L_{LP}$ - log $L_{PL}$). For all the epochs in the sources PKS 1510$-$089, 3C 279 and CTA 102 and the epoch A of 3C 273, LP model best describes the spectra, while the spectra during epoch B of 3C 273 is well fit by the PL model.}
\label{table-3}
{\begin{tabular}{cccrrccrrrr}
     \hline
     \multicolumn{4}{c}{PL} & \multicolumn{5}{c}{LP}&\\ 
     \cmidrule(lr){2-5} \cmidrule(lr){6-10}
         Epochs  & $\Gamma_{PL}$ & Flux  & TS & $-$Log L  & $\alpha$ & $\beta$ & 
    Flux & TS & $-$Log L & TS$_{curve}$\\
     \hline
PKS 1510-089 &             &              &         &         &                &               &                 &         &          &        \\
  A          &-2.41$\pm$0.04&0.08$\pm$0.006& 1502.89 & 32666.94 & 2.35$\pm$0.03 & 0.10$\pm$ 0.02 &  2.89$\pm$0.08 &  6626.5 & 30287.04 & 4759.8 \\
  B          &-2.38$\pm$0.01&1.10$\pm$0.02& 4484.27 &  40710.69 & 2.31$\pm$0.04 & 0.05$\pm$ 0.02 & 2.30$\pm$0.03 & 6046.02 &  39865.06 & 1691.24 \\
  C          &-2.29$\pm$0.05&1.05$\pm$0.03&  2143.16 & 26980.49 & 2.24$\pm$0.06 & 0.06$\pm$ 0.04 &  1.16$\pm$0.01 &   2149.74 & 26918.59 &  123.8 \\
  D          &-2.46$\pm$0.01&0.36$\pm$0.007&  1396.18 & 71986.58 & 2.40$\pm$0.05 & 0.009$\pm$ 0.007 &  0.38$\pm$0.009 &   1383.28 & 71943.76 &  85.64 \\
  E          &-2.35$\pm$0.02&0.10$\pm$0.006& 3632.06 & 41998.12 & 2.21$\pm$0.04 & 0.05$\pm$ 0.02 &  3.00$\pm$0.02 &  12254.5 & 37916.38 & 8163.48 \\
  F          &-2.26$\pm$0.02&0.44$\pm$0.008& 7069.12 & 33739.65 & 2.18$\pm$0.02 & 0.06$\pm$ 0.01 &  3.05$\pm$0.05 &  11891.14 & 31131.89 & 5215.52 \\
              \hline  
3C 273       &             &              &         &         &                &               &                 &         &          &        \\
 A           &-2.48$\pm$0.04&0.03$\pm$0.003& 384.26 & 18366.24 & 2.45$\pm$0.06 & 0.02$\pm$ 0.04 &  0.49$\pm$0.01 &  1965.47 & 17796.95 & 1138.58 \\
 B           &-2.91$\pm$ 0.10 & 0.005$\pm$1.13& 22.10 &  53603.31 & 2.91$\pm$0.11 & 0.13$\pm$ 0.16 & 0.46$\pm$0.02 & -328.63 & 53791.17 & -375.72 \\
     \hline     
3C 279       &             &              &         &         &                &               &                 &         &          &        \\
 A           &-2.42$\pm$0.03&0.02$\pm$0.003& 193.899 & 61574.51 & 2.46$\pm$0.07 & 0.06$\pm$ 0.01 &  0.16$\pm$0.004 &  341.966 & 61472.92 & 203.18 \\
 B           &-2.28$\pm$ 0.04 & 0.29$\pm$0.003& 5059.59 &  28075.17 & 2.13$\pm$0.05 & 0.09$\pm$ 0.02 & 0.61$\pm$0.03 & 6630.85 &  27337.68 & 1474.98 \\
 C           &-2.22$\pm$0.00&0.60$\pm$0.01&  14113.31 & 41165.00 & 2.05$\pm$0.03 & 0.10$\pm$ 0.01 &  4.09$\pm$0.001 &   25625.3 & 37399.62 &  7530.76 \\
 D           &-2.28$\pm$0.03&0.16$\pm$0.003&  2426.62 & 24463.13 & 2.14$\pm$0.04 & 0.08$\pm$ 0.02 &  1.44$\pm$0.01 &   5391.43 & 23282.19 &  2361.88 \\
     \hline
CTA 102      &             &              &         &         &                &               &                 &         &          &        \\
 A           &-2.49$\pm$0.00&0.002$\pm$0.91& 26.02 & 57614.41 & 2.58$\pm$0.06 & 0.02$\pm$ 0.05 &  0.16$\pm$0.004 &  565.75 & 57358.15 & 512.52 \\
 B           &-2.12$\pm$ 0.004 & 0.36$\pm$0.008& 10266.7 &  35236.65 & 1.84$\pm$0.03 & 0.09$\pm$ 0.01 & 0.63$\pm$0.02 & 14952.58 & 35028.19 & 416.92 \\
     \hline          
     \end{tabular}}
\end{table*}

\subsection{Spectral energy distribution modelling}

The sources studied here showed various characteristics in their optical and
$\gamma$-ray flux variations. There are instances when (a) optical
and $\gamma$-ray flux variations are correlated, (b) there is an optical
flare without a $\gamma$-ray counterpart and (c) there is a $\gamma$-ray flare
without an optical counterpart. To further characterize the nature
of the sources during the various epochs, we constructed their broad band 
SED during these epochs and studied them using simple one zone leptonic 
emission model. To obtain the SEDs in UV, optical and IR, all photometric 
measurements during each epoch were averaged filter wise to get one 
photometric point per epoch. In the
case of X-ray and $\gamma$-ray bands, all the data in each epoch were
used to generate the average X-ray and $\gamma$-ray spectra.

\begin{figure*}
\vspace*{-0.5cm}
\vbox{
\includegraphics[scale=0.56]{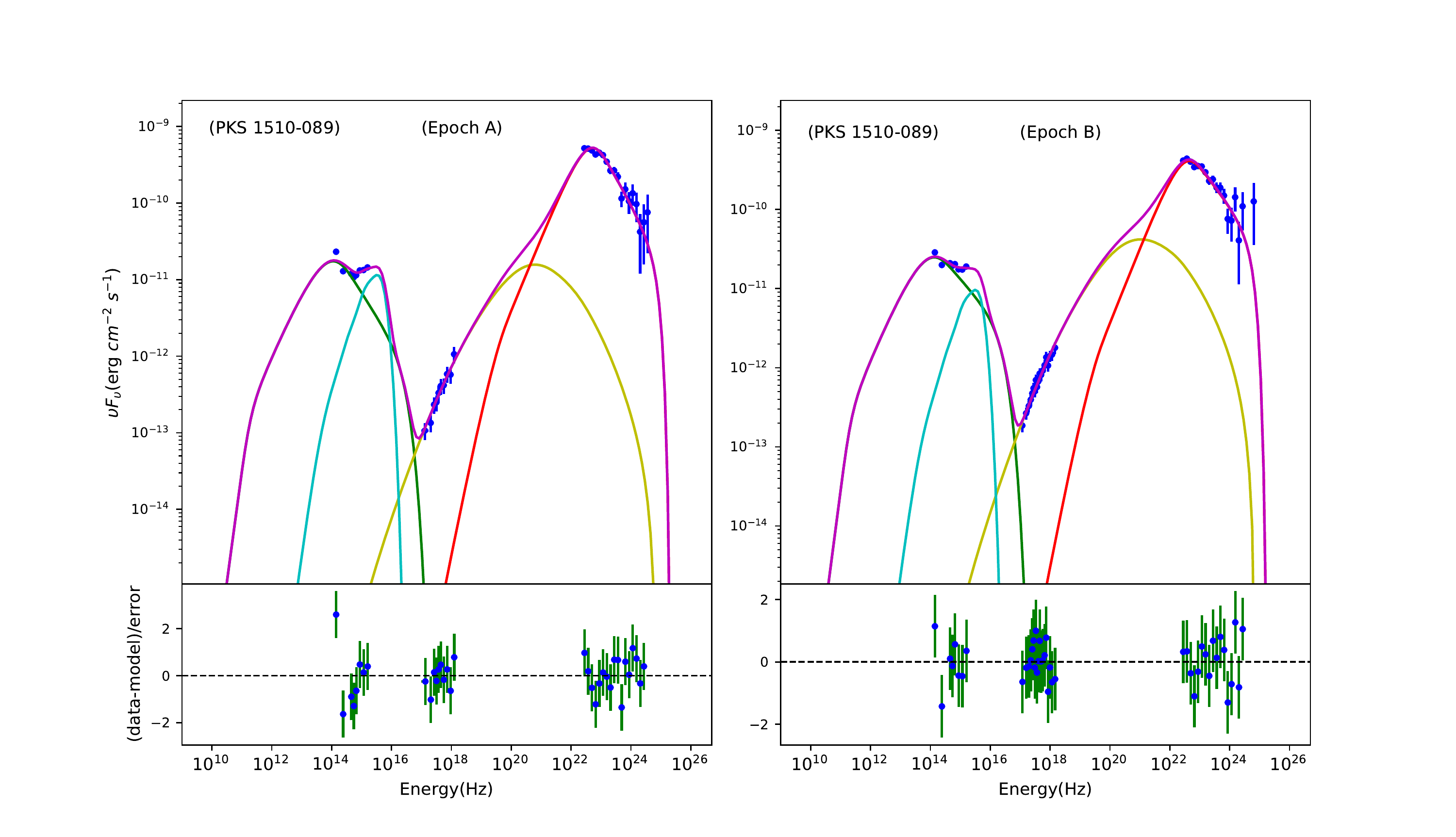}
\includegraphics[scale=0.56]{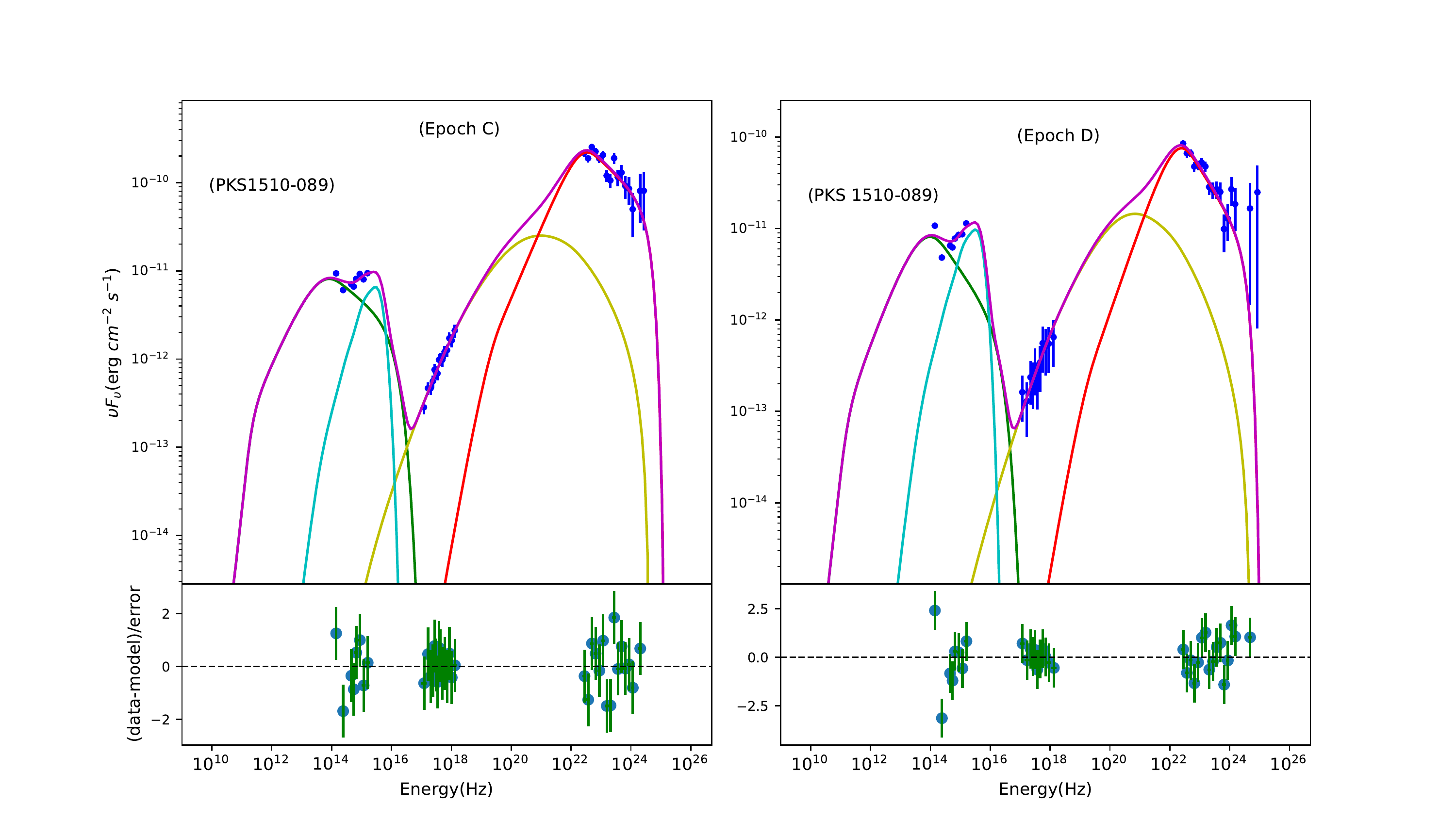}
     }
\vspace*{-0.6cm}\caption{Observed broad band spectral energy distribution 
for the source PKS 1510$-$089 along with model fits for the epochs A,B C and D. 
The various components are the synchrotron emission (green line), the SSC 
process (yellow line), the EC process (red line) and the accretion disk 
component (cyan line). The magenta line is the sum of all the components. The 
bottom panel in each SED residuals obtained by model fits to the observed data 
points obtained from XSPEC.}
\label{figure-12}
\end{figure*}

\begin{figure*}
\vspace*{-0.5cm}
\includegraphics[scale=0.56]{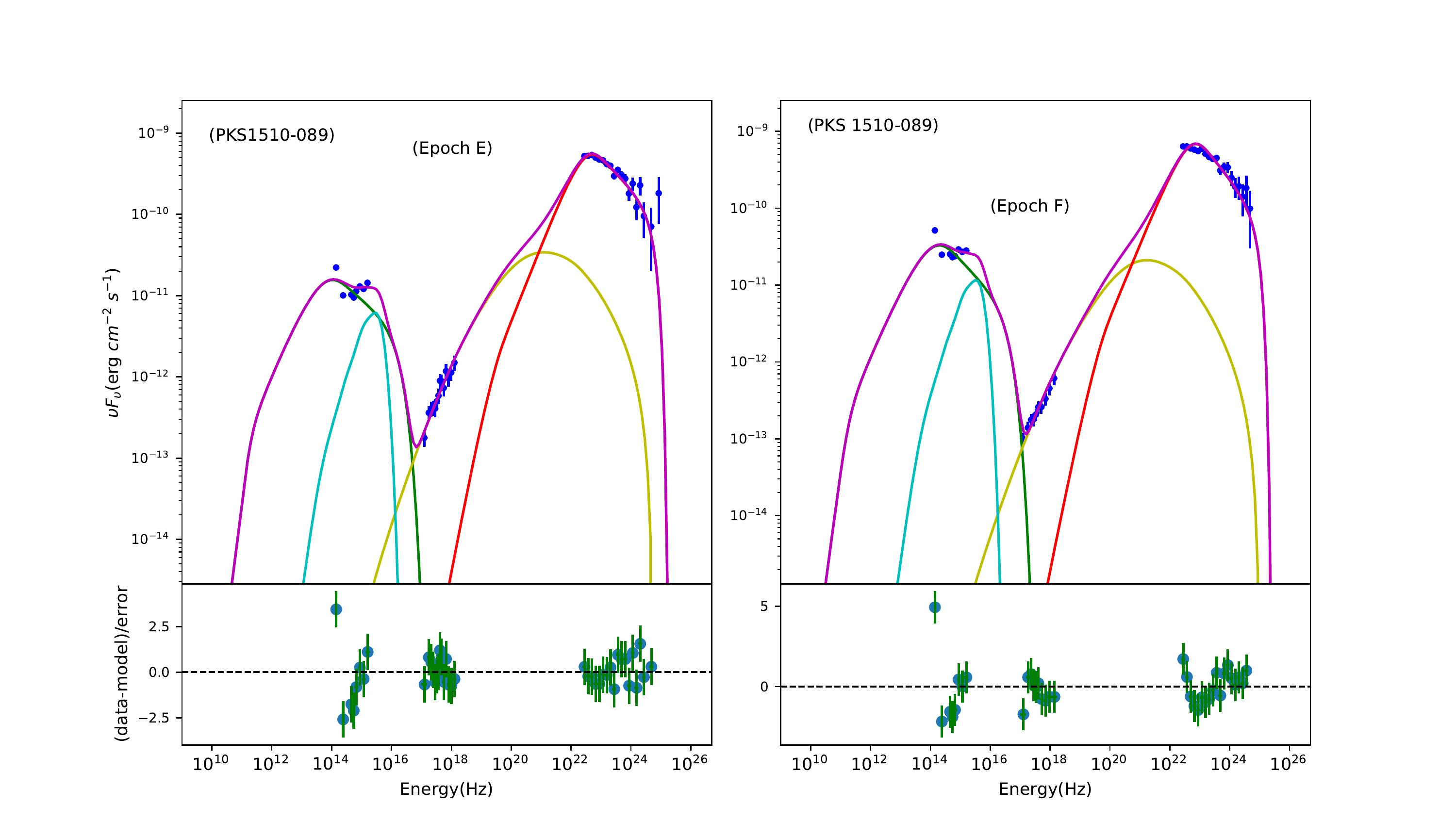}
\vspace*{-0.5cm}\caption{Model fits to the broad band SED during epochs E and F 
for the source PKS 1510$-$089. The lines and symbols are as in Fig. \ref{figure-12}}
\label{figure-13}
\end{figure*}

\begin{figure*}
\vspace*{-0.5cm}
\includegraphics[scale=0.56]{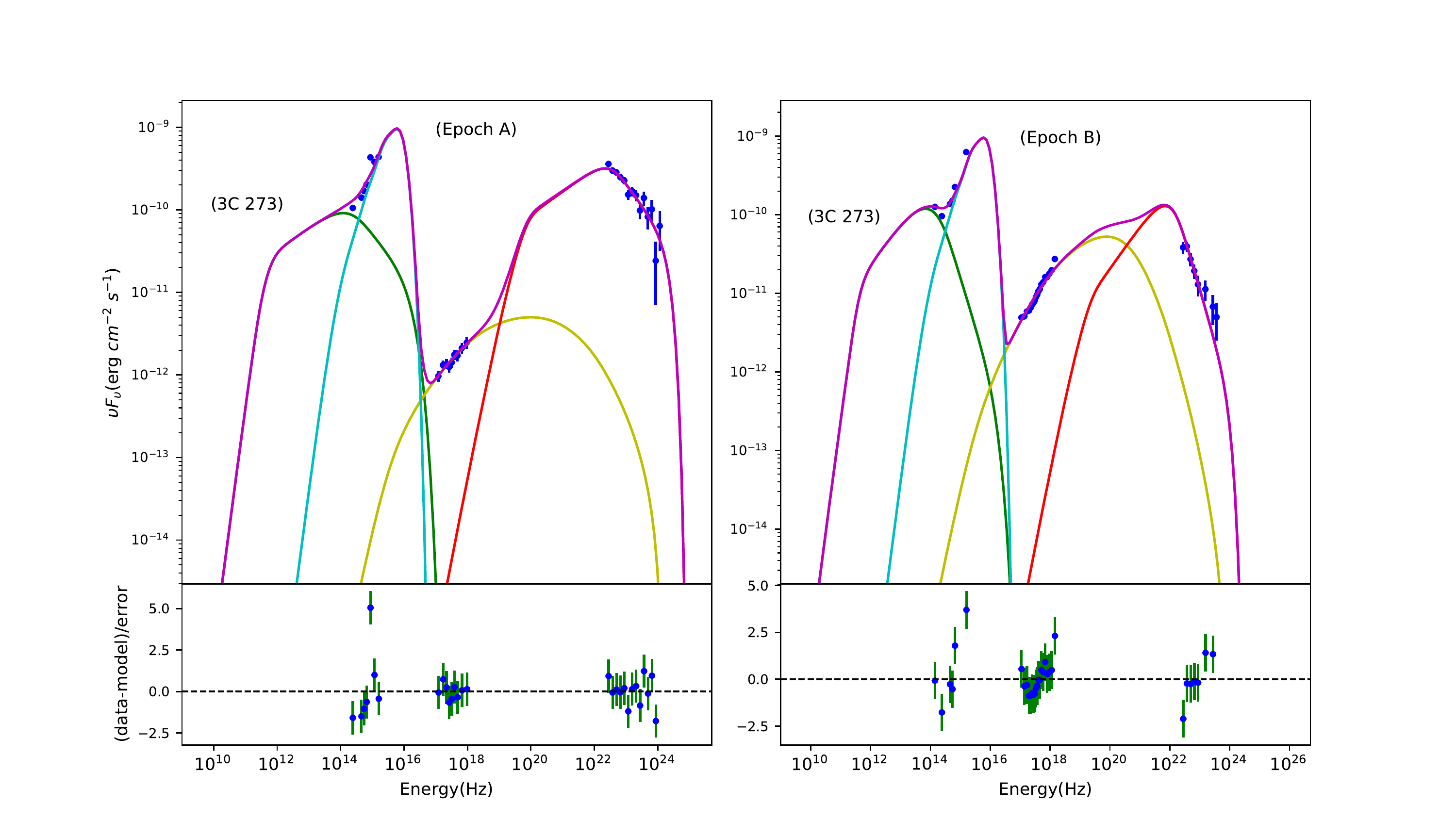}
\vspace*{-0.5cm}\caption{Model fits to the broad band SED during epochs A and B 
for source 3C 273. The lines and symbols are as in Fig. \ref{figure-12}}
\label{figure-14}
\end{figure*}

\begin{figure*}
\vspace*{-0.5cm}
\vbox{
\includegraphics[scale=0.56]{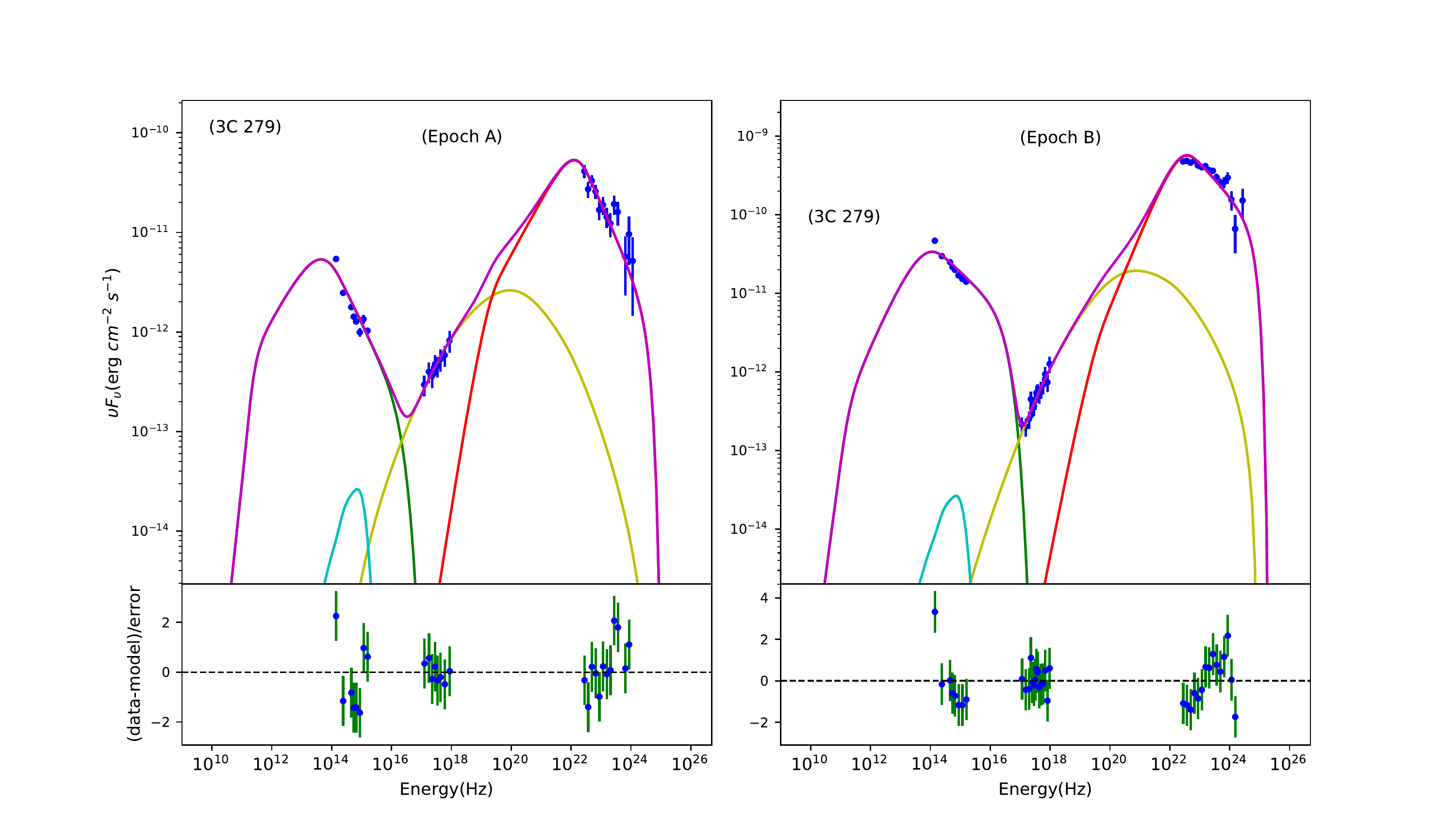}
\includegraphics[scale=0.56]{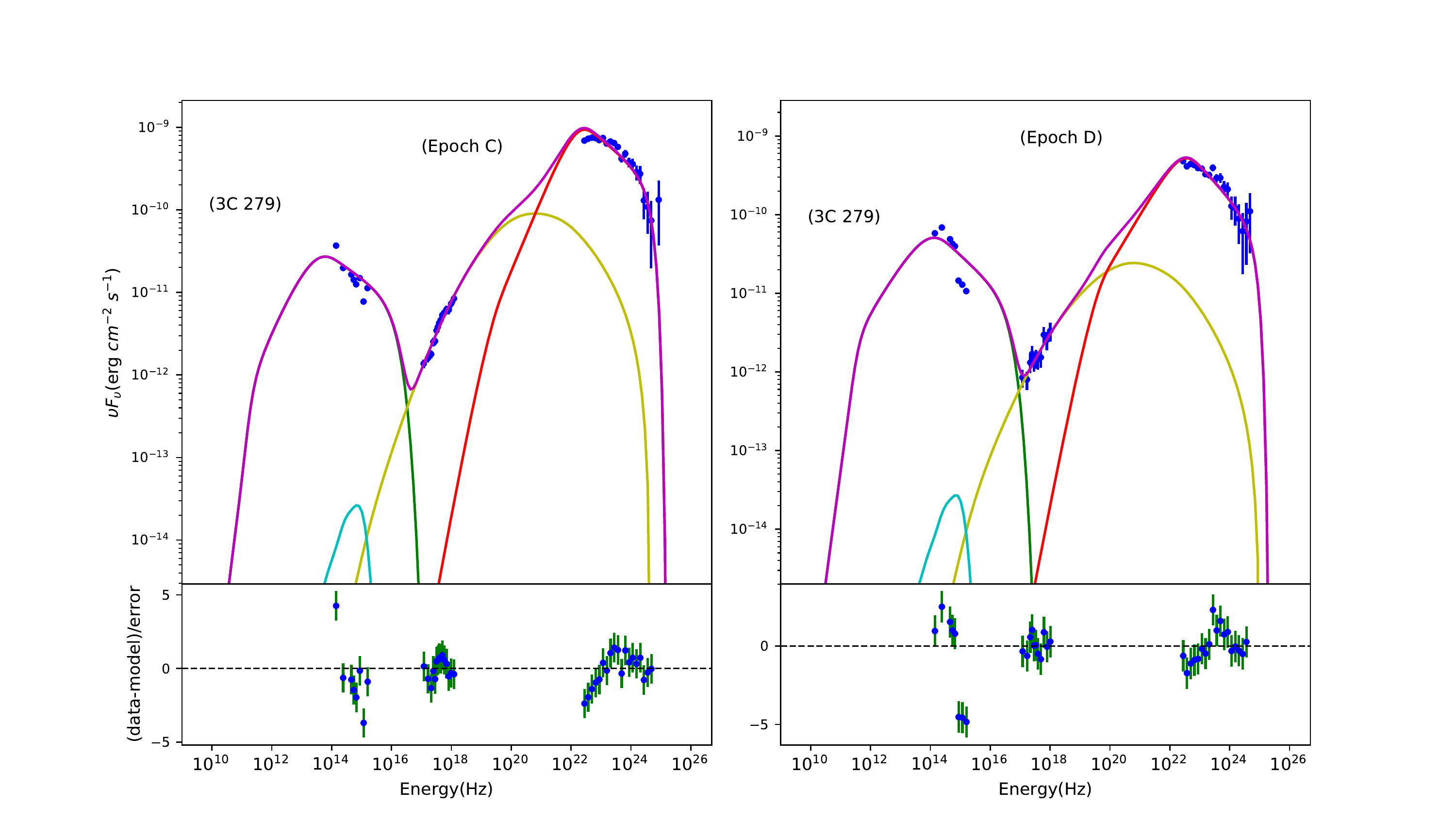}
     }
\vspace*{-0.5cm}\caption{Model fits to the broad band SED during epochs A, B, C and D for the 
source 3C 279. The lines and symbols are as in Fig. \ref{figure-12}}
\label{figure-15}
\end{figure*}

\begin{figure*}
\vspace*{-0.5cm}
\includegraphics[scale=0.56]{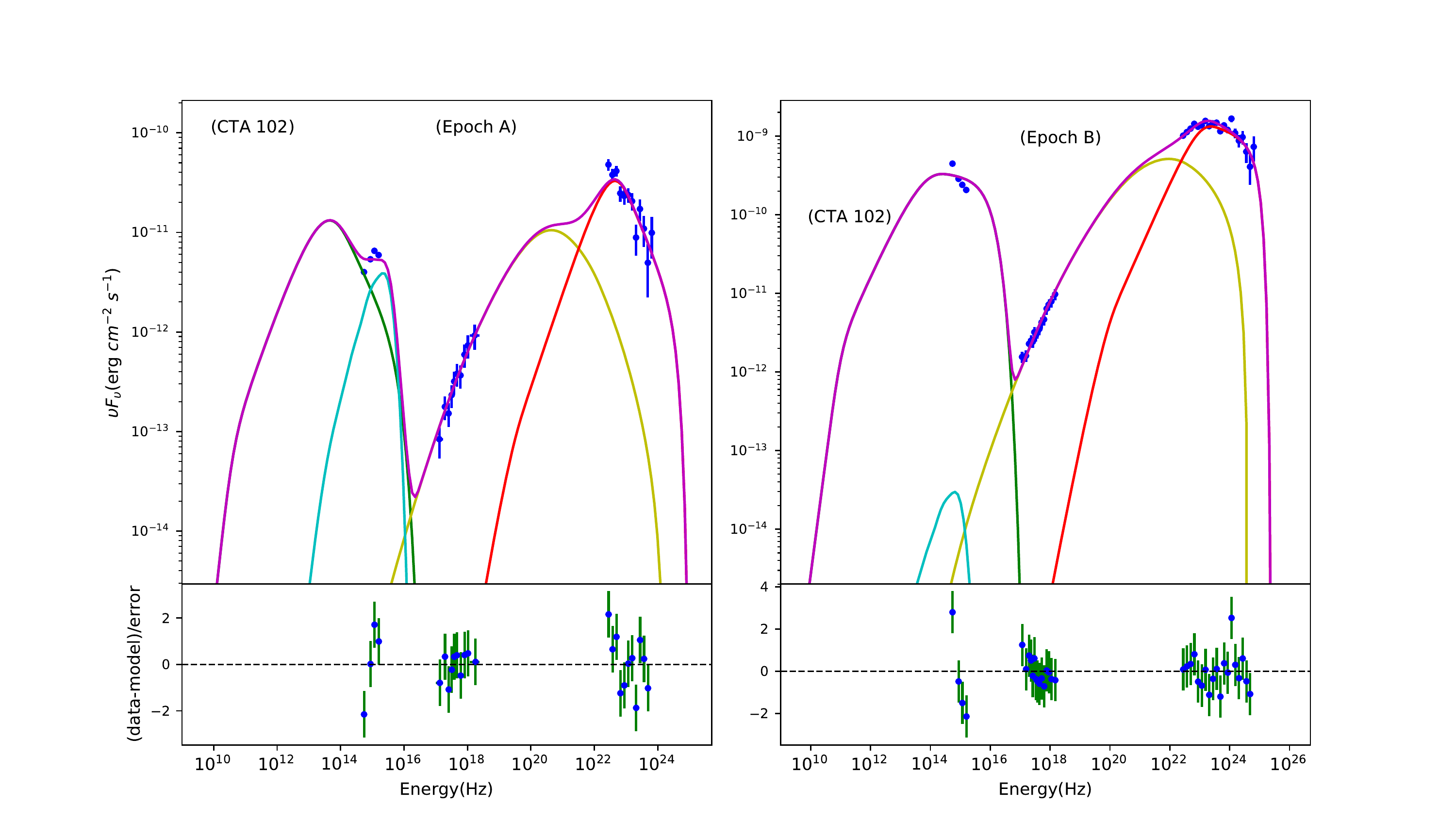}
\vspace*{-0.5cm}\caption{Model fits to the broad band SED during epoch A and B for the 
source CTA 102. The lines and symbols are as in Fig. \ref{figure-12}}
\label{figure-16}
\end{figure*}

Modelling the time averaged spectral energy distribution
during various epochs under different emission mechanisms 
can help us to understand
the physical conditions in the source. The $\gamma$-ray emission 
from FSRQs is generally interpreted as the external Compton scattering 
of thermal IR photons \citep{2019MNRAS.484.3168S,2019MNRAS.486.1781R} 
and hence the broad band SEDs are 
modeled using synchrotron, synchrotron self Compton and external
Compton emission processes. The details of the model are given in 
\cite{2018RAA....18...35S} and the best fit physical parameters
of the sources are obtained by fitting the broad band SED using 
$\chi^2$ minimization technique. We added 12\% systematics to the data in order
to account for the emission model related uncertainties. There
are twelve free parameters in our model, of which six parameters 
govern the electron energy distribution, namely electron energy index before 
the break (p), electron energy index after the break (q), the break Lorentz factor ($\gamma_b$), minimum Lorentz factor of the electron distribution ($\gamma_{min}$), the maximum Lorentz factor of the electron distribution ($\gamma_{max}$) and the electron
energy density ($U_e$). The other six parameters in the model are
the magnetic field (B), size of the emission region (R), Lorentz factor ($\Gamma$), 
jet viewing angle ($\theta$), external photon field
temperature (T) and the fraction of the external photons that take part in the
EC process (f). In order to investigate the different flaring behaviour 
between optical and $\gamma$-ray, firstly we fitted the quiescent epoch to obtain 
the parameters for all the sources. From the observed SED, we could 
obtain the high and low energy spectral indices, the synchrotron flux in the
optical, the SSC and EC fluxes in the X-ray and $\gamma$-ray energies respectively. 
Consistently for the model fit, we chose five free parameters namely p, q, $U_{e}$, 
$\Gamma$ and B while the other parameters were frozen to typical values. The adopted
values of the seven frozen parameters for the four sources are given in Table \ref{table-4}.
In Table \ref{table-5} and Figure\ref{figure-12} to Figure \ref{figure-16}, we summarize 
the 
results of the fitting. From the residuals given in the 
bottom panel of the Figures for some sources, it is clear the optical data is steeper and
deviate significantly from the best fit model. This suggests the presence of  
more than one emission component at optical/UV energies.\\ 
To account for the deviation of the model from the optical
spectra, we modified the model by including the emission from
the accretion disk.  The thermal emission from the disk is
decided by two parameters, namely the central black hole mass and the mass 
accretion rate \citep{1973A&A....24..337S,2008ApJ...676..351J}. The mass of the 
black hole is obtained from \cite{2018ApJS..235...39C} 
and the accretion rate is fitted to reproduce the optical spectra. This
procedure significantly improved the resultant $\chi^2$ and the best fit
parameters are given in Table \ref{table-5}. The model spectrum along with the 
observed data are given in Fig. \ref{figure-12} to \ref{figure-16} . Through this 
exercise we also demonstrate the capability to extract the accretion disk component 
from the broadband SED through a realistic spectral modelling involving different emission mechanisms.

\section{Results and Discussion}
\subsection{$\gamma$-ray spectra}
The high energy $\gamma$-ray spectra of FSRQs and low synchrotron peaked BL Lacs deviate
from the power law behaviour and are phenomenologically better represented either
as a broken power law (BPL) or a LP model. Such departures from simple 
PL fits noted  as a common feature in FSRQs firstly in the early observations
from {\it Fermi}-LAT \citep{2010ApJ...710.1271A} are now observed in the high energy spectra of
several FSRQs \citep{2014MNRAS.441.3591H,2015ApJ...803...15P,2019MNRAS.486.1781R,
2020A&A...635A..25S}. The cause of the spectral curvature seen in the
$\gamma$-ray spectra from {\it Fermi}-LAT is still not known. Several scenarios, both
intrinsic and extrinsic origins are proposed in the literature to explain the break in
the $\gamma$-ray spectrum of FSRQs.

One of the causes could be due to the attenuation of $\gamma$-rays by 
photon-photon pair production within the BLR due to HeII recombination and HI 
recombination. In this scenario termed as the double absorber model \citep{2010ApJ...717L.118P}, one expects to
see a break around 4$-$7 GeV and another break around 19.2$-$30 GeV. Such an
observation would imply absorption of $\gamma$-rays by BLR photons and the $\gamma$-ray
production site must lie within the BLR. However, observations do not support the
double absorber model \citep{2012ApJ...761....2H}. 
Alternatively, the break in the GeV spectra of FSRQs can happen by Klein-Nishina effect on the inverse Compton scattering of BLR photons by relativistic
jet electrons with a curved distribution \citep{2013ApJ...771L...4C}. But, from an analysis
of the $\gamma$-ray spectra of a large number of blazars, \cite{2018MNRAS.477.4749C} 
found that in FSRQs, the observed $\gamma$-ray spectra is not by IC scattering of BLR photons
and the $\gamma$-ray emission site lies outside the BLR.

\begin{figure}
\vbox{
\includegraphics[scale=0.57]{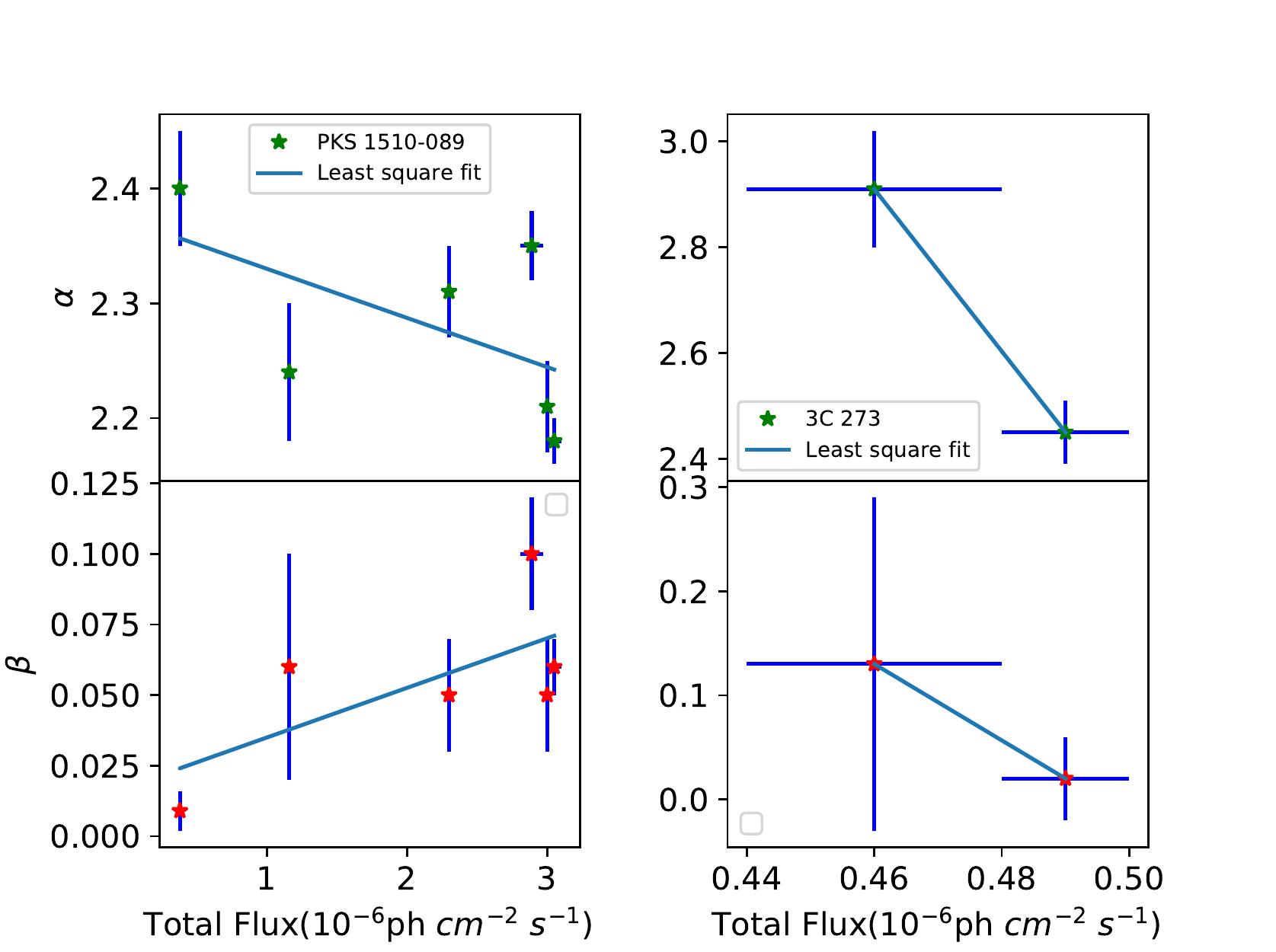}
\includegraphics[scale=0.57]{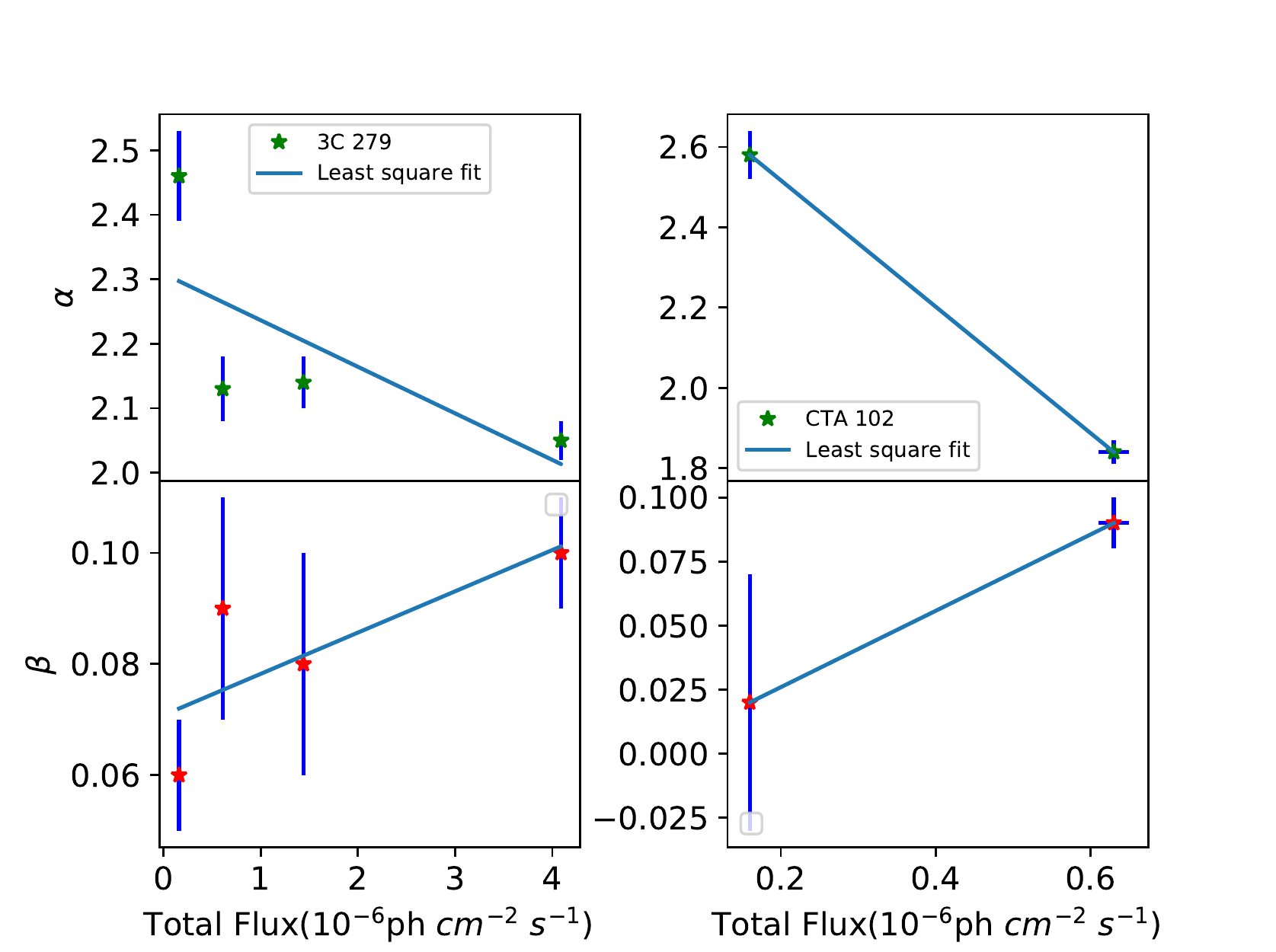}
}
\caption{Variations of the parameters $\alpha$ and $\beta$ with flux for
the sources PKS 1510$-$089 (top left), 3C 273 (top right), 3C 279 (bottom
left) and CTA 102 (bottom right).}
\label{figure-17}
\end{figure}

Apart from the above, the break in the $\gamma$-ray spectra of FSRQs can also happen due to
intrinsic effects because of the electrons in the relativistic jets of these sources
either having a cut-off in their energy distribution or a log-parabola energy
distribution. In this work, the SEDs of all the sources in the different epochs are well modelled by
IC scattering of the photons from the obscuring torus, and the $\gamma$-ray emission region
lies outside the BLR where IC takes place in the Thomson regime. The results
of the $\gamma$-ray spectral analysis carried out on all the epochs in the objects
and reported in Table \ref{table-3}. We note that in the 3FGL catalog
(https://www.ssdc.asi.it/fermi3fgl/), the $\gamma$-ray spectra of all the sources studied here are better described by the LP model than the PL model. In this work, the $\gamma$-ray spectra of all the sources at all the epochs are better described by the LP model except for epoch B of 3C 273, which is well fit by the PL model. The parameters $\alpha$ and $\beta$ in the LP model fits to the data carry very important information on the characteristics of the $\gamma$-ray spectra. In this model, $\alpha$ gives the slope of the spectra and $\beta$ is a measure of the curvature in the spectra. A smaller value of 
$\alpha$ and $\beta$ implies a harder spectrum with a mild curvature. Any 
changes in the value of $\alpha$ and $\beta$ during
different epochs is a measure of the changes in the $\gamma$-ray spectral shape. 
The dependence of $\alpha$ and $\beta$ values against the fluxes of the sources are given in 
Fig. \ref{figure-17}. For all the sources we found the spectra to harden with increasing
flux. We found decreasing as well as increasing trend of $\beta$ with flux. The 
variation in the $\gamma$-ray spectral shape can be associated with the shift 
in IC peak frequency. This is evident from the results of our SED analysis. 
Our model fits to the observed SED also gives the IC peak frequency (see 
Table \ref{table-5}). Analysis of the IC peak indicates that as the IC peak shifts towards
lower energies, the spectrum is harder and the curvature ($\beta$) is sharper which too
demonstrates that the $\gamma$-ray spectral variation is closely related
to the changes in the IC peak. Alternatively, $\gamma$-ray spectral 
variation can also be attributed to the changes in the location of the 
$\gamma$-ray emission region during different activity states of the sources \citep{2016MNRAS.458..354C}. Besides, since the $\gamma$-ray emission in FSRQs is due to EC scattering of the external target photons, the $\gamma$-ray peak energy will depend on the dominant external photon frequency. If the target photon field is the IR emission from the dusty torus, then the temperature of the dust emission will depend on the location of the emission region from the central black hole \citep{2014ApJ...782...82D,2009MNRAS.397..985G}.
\subsection{Connection between optical and GeV flux variation}
\begin{figure}
\includegraphics[scale=0.6]{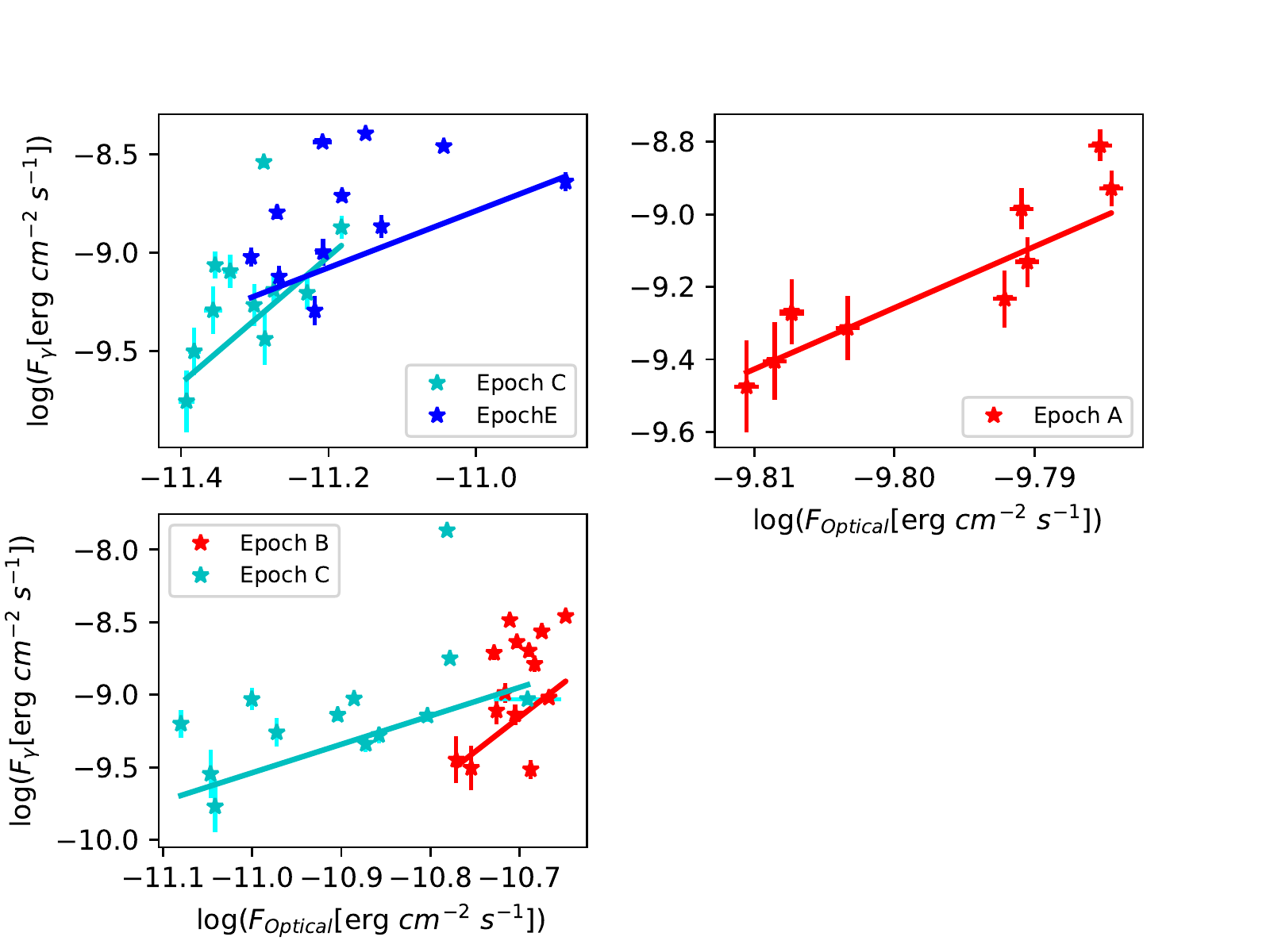}
\caption{Optical flux v/s $\gamma$-ray flux for the sources PKS 1510$-$089 (top left), 
3C 273 (top right) and 3C 279 (bottom left) respectively.}
\label{figure-18}
\end{figure}

The capability of {\it Fermi} to scan the sky once in three hours and supporting ground based
monitoring observations in the optical band has enabled one to study close correlations
between flux variations in the GeV band and other low energy bands. From multiband
observations of the blazar 3C 454.3 over a period of about 5 months, 
\cite{2009ApJ...697L..81B}
found close correlation between the optical and GeV band flux variations. This argues for
co-spatiality of the optical and GeV emission regions. This correlation is also easily
understood in the one zone leptonic emission model, wherein relativistic electrons in the jet produce
optical emission by synchrotron process, and the same relativistic electrons
produce $\gamma$-ray emission by inverse Compton process. However, analysis of the
same source by \cite{2019MNRAS.486.1781R} noticed that the optical and GeV flux
variations are not correlated at all times. Such mismatch between optical and
GeV flux variations are also known in few other blazars such as PKS 0208$-$512, 
\citep{2013ApJ...763L..11C}, PKS 0454$-$234, S4 1849+67, BZQ J0850$-$1213, OP 313 
\citep{2014ApJ...797..137C},
PKS 2142$-$75 \citep{2013ApJ...779..174D}, PKS 1510$-$089 
\citep{2015ApJ...804..111M} and PKS 2155$-$304 \citep{2019Galax...7...21W}.
From the analysis of multi-band light curves
of the sources, we found instances where the optical and $\gamma$-ray flux variations 
are closely correlated,
cases where there are optical flares without $\gamma$-ray counterpart and instances 
when there are $\gamma$-ray flares without optical counterparts. 
Thus, it is evident that the correlations between the optical and GeV flux 
variations in {\it Fermi} blazars are complex. Recently,
from correlation analysis between the optical and $\gamma$-ray light curves of 178 blazars, 
\cite{2019ApJ...880...32L} found that statistically about 50\% of their optical 
flares have no GeV counterparts and this fraction is less in the case of $\gamma$-ray flares, 
i.e., about 20\% of $\gamma$-ray flares have no optical counterparts. 
While in the leptonic scenario a close correlation between optical and GeV variations is expected, the results found in this work as well as the other recent 
results
by \cite{2019MNRAS.486.1781R}  and \cite{2019ApJ...880...32L} indicate that 
correlated variability analysis between
the optical and GeV bands may also not be definitive in constraining the 
leptonic v/s hadronic scenario
for the high energy emission process in blazars.
Most of the correlation studies between different energy bands of the blazars 
indicate positive correlation. But there are exceptions and cases of anticorrelation 
are also found for some sources \citep{2013ApJ...763L..11C,2014ApJ...797..137C,2013ApJ...779..174D,2015ApJ...804..111M,2019MNRAS.486.1781R}. 
We found various behaviours between optical and $\gamma$-ray energy bands for our 
selected sample of sources. We looked for a correlation between the optical (V-band) and $\gamma$-rays for all the epochs considered here. Fig. \ref{figure-18} shows only the epochs where the correlation is significant at the 90\% level. We converted the $\gamma$-ray fluxes from ph $cm^{-2}$ $s^{-1}$ to erg $cm^{-2}$ $s^{-1}$ at 100 MeV \citep{2014ApJ...786..109S} to match the optical flux units. The results of the fit are given in Table \ref{table-7}. The fit takes into account the uncertainty in both optical and $\gamma$-rays. For the source PKS 1510$-$089 during epochs B and F 
the optical and $\gamma$-ray flares are correlated. During this epoch $\Gamma$ is larger than that 
of the quiescent period. This has given rise to increased flare in optical and $\gamma$-rays. The difference
in the amplitude of variations between optical and $\gamma$-ray flares during epochs B and F must
be due to a combination of $\Gamma$ and B. For epochs A, C and E the magnetic field is lower than the 
quiescent period by a factor of 1.2 - 1.5. This has led to decreased optical variations. At the same time
$\Gamma$ has increased from 1.1 - 1.7 times the quiescent period leading to increased $\gamma$-ray flare,
but no corresponding optical flare (see Fig. \ref{figure-2} and Fig. \ref{figure-3}).

In the source 3C 273, using our criteria, we were able to identify one quiescent 
period and one flaring period. At epoch A, the bulk Lorentz factor increased compared to the quiescent state B, whether the magnetic field is nearly the same (Table \ref{table-5}). It is natural to expect increased optical and $\gamma$-ray flares during epoch A, but we found a $\gamma$-ray flare without an optical counterpart. This absence in optical flux 
variations might be due to sub-dominant optical synchrotron emission compared to the 
prominent accretion disk emission. The prominent accretion disk component is
very well evident in the broad band SED both in the quiescent and active 
states (Fig. \ref{figure-14}).

In the source 3C 279, we identified four epochs, A,B, C and D of which during 
epoch A, the source was quiescent while it was active during the other epochs. 
During epoch D, magnetic field is about 1.5 times larger than the 
quiescent period leading to larger optical synchrotron emission. At the same 
epoch, $\Gamma$ has increased from about 7 to 12. This explains the increased 
$\gamma$-ray and optical flaring in epoch D. During epochs B and C, $\Gamma$ has increased 
relative to the quiescent state giving rise to larger $\gamma$-ray fluctuations. During 
epoch B, magnetic field is marginally larger than the quiescent period, while the 
particle density is lower than the quiescent period. However, in epoch C, the magnetic field and particle density is lower and
higher respectively than the quiescent period. The interplay between low magnetic field and high particle density
and vice versa could lead to lower optical variations. This could be the reason for $\gamma$-ray flares
without optical counterparts in epochs B and C in 3C 279. The low energy peak of the SED during all epochs in 3C 279 is dominated by synchrotron emission from the relativistic jet.

In CTA 102, we found one flaring epoch when the optical and $\gamma$-ray seems to be correlated. Many short term flares
with optical and $\gamma$-ray counterparts are seen during this epoch. For SED analysis we considered only 10 days
due to the availability of $\gamma$-ray, X-ray and optical data points for SED 
modelling. During Epoch B, $\Gamma$ was nearly four times greater than the quiescent epoch. The magnetic field during epoch B and the quiescent period agree with each other within errors (see Table \ref{table-5}). This increase
in $\Gamma$ relative to the quiescent epoch is the cause of the increased $\gamma$-ray flare and optical flare during epoch B. Prominent
accretion disk component is visible in the SED during the quiescent phase, however, this is not evident
in the flaring epoch B (Fig. \ref{figure-16}). This is also reflected in the high accretion rate found during the quiescent
epoch A and a negligibly small accretion rate during epoch B.

\begin{figure}
\includegraphics[scale=0.6]{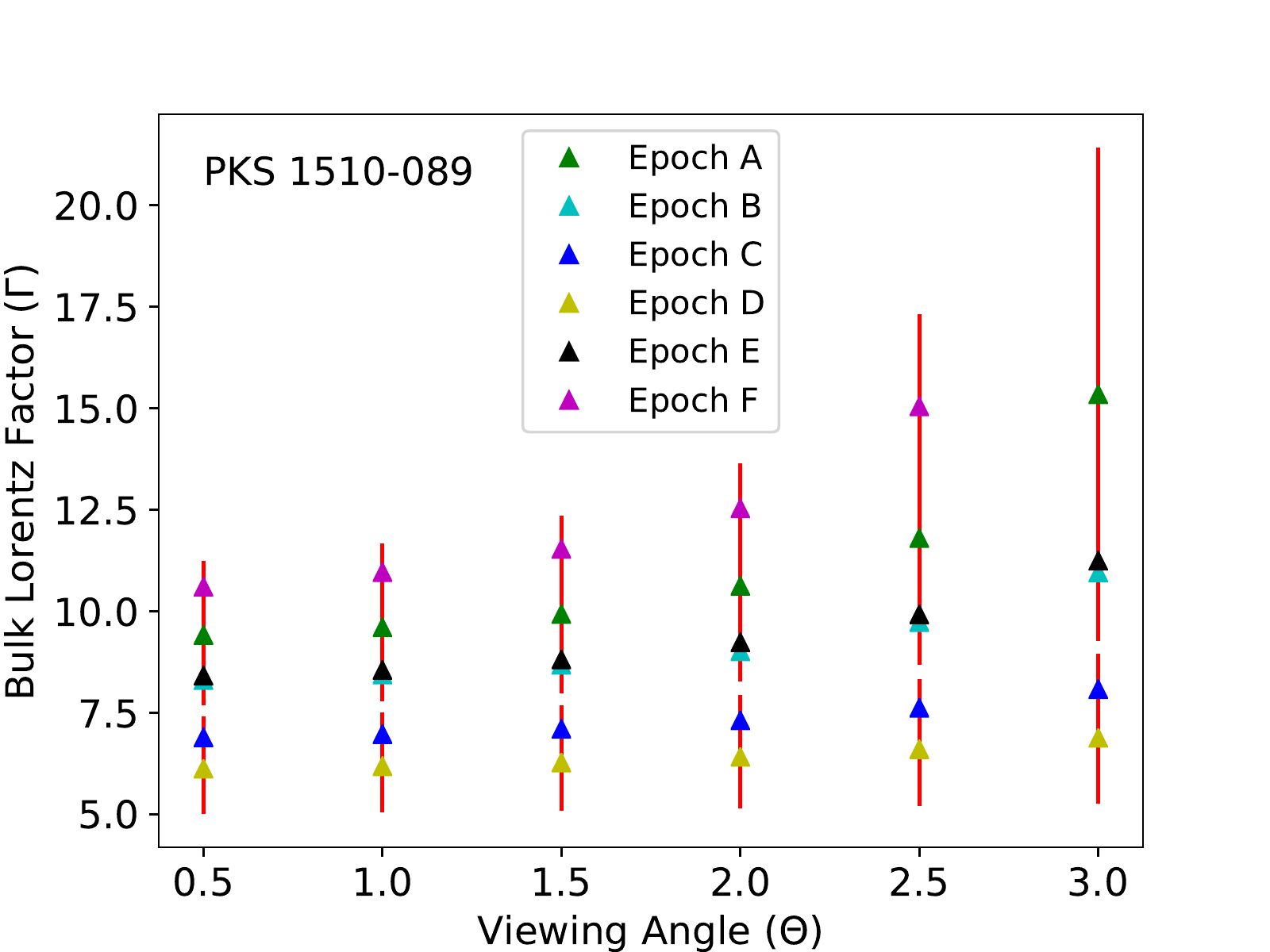}
\caption{Bulk Lorentz factor v/s viewing angle for the sources PKS 1510$-$089.}
\label{figure-19}
\end{figure}

In the SED analysis carried out in this work, the viewing angle is fixed 
at 2$^\circ$ for all the sources. Radio observations indicate that the average 
viewing angle of the sources studied here is lesser than 2$^\circ$ except 
for 3C 273, where it is around 6$^\circ$ \citep{2017ApJ...846...98J}. Similarly, according to 
\cite{2009A&A...494..527H} the average viewing angle is around 3.5$^\circ$ for the sources 
except CTA 102 for which it is 3.7$^\circ$. Thus the values of viewing angle available
in the literature of these source is generally low and not too different from the constant
value of 2$^\circ$ used for all the sources. However, to ascertain the effect
viewing angle can have on the $\Gamma$ obtained from SED analysis, we repeated the 
SED analysis for different values of $\theta$ such as 0.5, 1.0, 1.5, 2.5 and 3.0 degrees. A similar 
fit statistics was obtained for all cases with considerable increase in the $\Gamma$ suggesting a degeneracy. The plot between $\Gamma$ v/s $\theta$ for the source PKS 1510$-$089 is shown in the Figure \ref{figure-19}. We found that as $\theta$ increases, $\Gamma$ too increased. No significant changes were noticed in the other parameters, and the changes are consistent within the errors. The choice of viewing angle do not alter our conclusions since the Lorentz factor obtained for different epochs still follow the same trend. This trend is also found in other sources. To further verify the degeneracy between the bulk Lorentz factor and the viewing angle we continued the fitting procedure with the constant $\Gamma$, which is fixed at 15 and repeated the SED fitting for all the sources with the inclusion of the accretion disk component. The results of the SED fitting at constant $\Gamma$ are given in Table \ref{table-6}. We found minimal change in the parameters compared to their values given in Table \ref{table-5}. It shows that the adopted $\theta$ values directly impact the predicted $\Gamma$ or vice-versa and this is true for all the sources and at all epochs. Recent study from radio observations of the source S5 0716 + 714 by \cite{2020ApJ...893...68K} also shows that the variation in $\Gamma$ occurs due to the change in $\theta$. This indicates that the changes in the $\gamma$-ray flux states of a source is largely associated with the changes in the $\Gamma$ of the jet, as well as $\theta$. Though various 
physical processes have been proposed
to explain the uncorrelated optical and GeV flux variations in blazars in the 
literature, our analysis of the broad band SED of the four FSRQs studied in this 
work on different epochs is consistent with leptonic processes in the jets of 
these sources. 
\subsection{Optical spectral variations}
It is argued that FSRQs generally show a RWB trend \citep{2019ApJ...887..185S}, 
while BL Lacs show a BWB trend \citep{2019MNRAS.484.5633G}. 
If these two classes of blazars indeed show a distinct colour magnitude relation, it leads
to hypothesize that the jets are fundamentally different between FSRQs and BL Lacs.   
However, with the availability of more data it is now known that blazars show different types 
of colour variability in the optical - IR bands. 
Blazars are found to show BWB trend (e.g., \citealt{2009MNRAS.399.1357S}), RWB trend  
(e.g., \citealt{2019ApJ...887..185S}), both BWB and RBW trends \citep{2019MNRAS.486.1781R}
as well as weak/no spectral change with brightness \citep{2003A&A...402..151R}.
The dominance of the more variable red synchrotron emission over the less variable 
thermal emission from the accretion disk could lead to a RWB trend \citep{2019ApJ...887..185S}.
Alternatively, a BWB trend can happen due to increased variations at shorter wavelengths 
\citep{2009MNRAS.399.1357S}.
In the one zone leptonic emission model this can be explained via the injection of fresh
electrons with high energy leading to a BWB trend
\citep{1998A&A...333..452K,2002PASA...19..138M}. Alternatively, according to \cite{2007A&A...470..857P} and \cite{2004A&A...421..103V}
a BWB trend can also happen due to changes in the Doppler factor. 
We examined the colour
variations during all the epochs and considered a source to show colour variation 
only when the Spearman rank correlation coefficient is $>$ 0.5 as well as $<$ $-$0.5 
with the null hypothesis probability $<$ 0.05. The observed colour variation 
is significant at the 95\% level. In the case of PKS 1510$-$089 significant colour 
variations were observed during epoch A,E and F. While during epochs A and E, we found
a RWB trend, during epoch R, we found a BWB trend. The observed colour variation 
in the optical band has no direct relation to the presence or absence of correlated 
variation between optical and $\gamma$-ray bands. Our results clearly indicate that 
FSRQs show both BWB and RWB behaviours.

\begin{table}
\caption{Values of the parameters that were frozen during the model fits to the 
observed SEDs. Here, R is the size of the emission region in units of 10$^{15}$ cm, and the temperature T is in Kelvin.}
\label{table-4}
\addtolength{\tabcolsep}{-1.6pt}
\begin{tabular}{lcrrrrcl} \hline
Object        & R & $\gamma_{min}$    & $\gamma_{max}$   & $\gamma_b$  & T(K)  & f   \\ \hline
PKS 1510$-$089  & 7.9    &   40  &  2 $\times$ 10$^4$  & 1500 &  800 & 0.9   \\
3C 273          & 15.8   &   50  &  1 $\times$ 10$^4$  & 1200 &  800 & 0.9    \\
3C 279          & 15.8   &   40  &  2 $\times$ 10$^4$  & 1200 &  800 & 0.9     \\
CTA 102         & 100    &   50  &  2 $\times$ 10$^4$  & 2100 &  800 &  0.02    \\
\hline
\end{tabular}
\end{table}

\begin{table*}
\caption{Results of the broad band SED analysis of the sources at different epochs for the constant viewing angle $\theta$=$2^{\circ}$}.
\label{table-5}
\addtolength{\tabcolsep}{-1.2pt}
\begin{tabular}{llrccccrcc} \hline
      &       & Bulk Lorentz & Low energy     & High energy    & Eletron energy      & Magnetic      & Accretion &  IC peak  &           \\
Name  & Epoch &  factor      & particle index & particle index & density (cm$^{-3}$) & field (Gauss) & rate      &  (MeV)   &  $\chi^2$/dof \\
\hline
PKS 1510$-$089 & A (with out AD) & 10.29 $\pm$ 1.12    & 1.10 $\pm$ 0.52 & 3.85 $\pm$ 0.11 & 0.12 $\pm$ 0.04  & 0.97 $\pm$ 0.08  & ---              & 227  &  1.9 \\ 
             & A (with AD)     & 10.62 $\pm$ 1.20    & 1.10 $\pm$ 0.58  & 4.22 $\pm$ 0.16  & 0.12 $\pm$ 0.04  & 0.84 $\pm$ 0.07  & 1.48 $\pm$ 0.32  &   &  0.8 \\ 
             & B (with out AD) &  8.56 $\pm$ 0.65    & 1.10 $\pm$ 0.37  & 3.66 $\pm$ 0.09  & 0.22 $\pm$ 0.05  & 1.05 $\pm$ 0.06  & ---              & 198  & 0.8 \\
             & B (with AD)     &  9.01 $\pm$ 0.69    & 1.10 $\pm$ 0.38  & 3.86 $\pm$ 0.12  & 0.20 $\pm$ 0.05  & 0.98 $\pm$ 0.06  & 1.24 $\pm$ 0.48  &  &  0.5 \\
             & C (without AD)  &  5.72 $\pm$ 0.61    & 1.10 $\pm$ 0.35  & 3.04 $\pm$ 0.13  & 0.66 $\pm$ 0.19  & 0.66 $\pm$ 0.03  & ---              & 163  & 1.2 \\
             & C (with AD)     &  7.31 $\pm$ 0.63    & 1.39 $\pm$ 0.23  & 3.55 $\pm$ 0.13  & 0.37 $\pm$ 0.09  & 0.64 $\pm$ 0.05  & 0.85 $\pm$ 0.22  &  &  0.7 \\
             & D (without AD)  &  5.00 $\pm$ 1.31    & 1.10 $\pm$ 0.36  & 3.20 $\pm$ 0.13  & 0.32 $\pm$ 0.09  & 1.34 $\pm$ 0.08  & ---              & 116  & 2.4 \\
             & D (with AD)     &  6.41 $\pm$ 1.27    & 1.10 $\pm$ 1.26  & 3.95 $\pm$ 0.18  & 0.21 $\pm$ 0.15  & 0.98 $\pm$ 0.16  & 1.26 $\pm$ 0.22  &  &  1.2 \\
             & E (without AD)  &  8.62 $\pm$ 0.86    & 1.10 $\pm$ 0.51  & 3.45 $\pm$ 0.08  & 0.27 $\pm$ 0.09  & 0.72 $\pm$ 0.06  & ---              & 209  & 1.6 \\
             & E (with AD)     &  9.22 $\pm$ 0.96    & 1.13 $\pm$ 0.27  & 3.63 $\pm$ 0.10  & 0.24 $\pm$ 0.08  & 0.68 $\pm$ 0.06  & 0.79 $\pm$ 0.28  &      & 1.3 \\
             & F (without AD)  & 12.31 $\pm$ 1.19    & 1.10 $\pm$ 0.37  & 3.68 $\pm$ 0.09  & 0.08 $\pm$ 0.02  & 1.18 $\pm$ 0.07  & ---              & 333  & 2.2 \\
             & F (with AD)     & 12.53 $\pm$ 1.11    & 1.10 $\pm$ 0.33  & 3.81 $\pm$ 0.12  & 0.07 $\pm$ 0.02  & 1.12 $\pm$ 0.06  & 1.49 $\pm$ 0.67  &      & 1.9 \\ \hline
3C 273       & A (with AD)     & 9.42 $\pm$ 1.37     & 2.44 $\pm$ 0.27  & 3.94 $\pm$ 0.15  & 0.01 $\pm$ 0.004 & 2.21 $\pm$ 0.42  & 16.00 $\pm$ 4.13  & 121   & 1.9 \\
             & B (with AD)     & 5.00 $\pm$ 0.52 & 1.92 $\pm$ 0.12  & 5.48 $\pm$ 0.44  & 0.07 $\pm$ 0.009  & 2.33 $\pm$ 0.20  & 16.00 $\pm$ 2.72  & 32   &  1.5 \\ \hline
3C 279       & A (without AD)  &  7.13 $\pm$ 1.03    & 1.89 $\pm$ 0.46  & 4.27 $\pm$ 0.10  & 0.07 $\pm$ 0.02  & 1.05 $\pm$ 0.14  & ---              & 64   & 1.5 \\
             & A (with AD)     &  7.13 $\pm$ 1.09    & 1.89 $\pm$ 0.48  & 4.25 $\pm$ 0.20  & 0.07 $\pm$ 0.02  & 1.05 $\pm$ 0.15  & 0.01 $\pm$ 0.15  &      & 1.7 \\
             & B (without AD)  & 11.22 $\pm$ 1.10    & 1.15 $\pm$ 0.26  & 3.71 $\pm$ 0.08  & 0.04 $\pm$ 0.01  & 1.16 $\pm$ 0.09  & ---              & 177  & 1.1 \\
             & B (with AD)     & 11.22 $\pm$ 1.12    & 1.15 $\pm$ 0.26  & 3.71 $\pm$ 0.10  & 0.04 $\pm$ 0.01  & 1.16 $\pm$ 0.09  & 0.01 $\pm$ 1.18  &      & 1.2 \\            
             & C (without AD)  &  8.80 $\pm$ 0.04    & 1.24 $\pm$ 0.20  & 3.51 $\pm$ 0.07  & 0.22 $\pm$ 0.03  & 0.66 $\pm$ 0.04  & ---              & 147  & 1.8 \\
             & C (with AD)     &  8.80 $\pm$ 0.60    & 1.24 $\pm$ 0.21  & 3.51 $\pm$ 0.09  & 0.22 $\pm$ 0.04  & 0.66 $\pm$ 0.04  & 0.01 $\pm$ 1.86  &      & 1.9 \\
             & D (without AD)  & 11.75 $\pm$ 1.53    & 1.72 $\pm$ 0.25  & 3.70 $\pm$ 0.01  & 0.05 $\pm$ 0.01  & 1.52 $\pm$ 0.13  & ---              & 154  & 3.0 \\
             & D (with AD)     & 11.75 $\pm$ 1.50    & 1.72 $\pm$ 0.26  & 3.70 $\pm$ 0.11  & 0.05 $\pm$ 0.01  & 1.52 $\pm$ 0.18  & 0.01 $\pm$ 4.92  &      & 3.1 \\ \hline
CTA 102      & A (without AD)  & 7.98 $\pm$ 1.50 & 1.10 $\pm$ 0.64  & 3.84 $\pm$ 0.46  & 0.006 $\pm$ 0.004  & 0.39 $\pm$ 0.10  & ---              & 196  & 2.4 \\
             & A (with AD)     & 8.42 $\pm$ 1.19 & 1.10 $\pm$ 0.70  & 4.35 $\pm$ 0.38 & 0.006 $\pm$ 0.003  & 0.32 $\pm$ 0.07  & 10.00 $\pm$ 6.09 &    & 1.4 \\
             & B (without AD)  & 32.17 $\pm$ 13.61 & 1.29 $\pm$ 0.08 & 3.17 $\pm$ 0.15  & 0.007 $\pm$ 0.001 & 0.42 $\pm$ 0.03  & ---              & 1295 & 1.0 \\
             & B (with AD)     & 36.44 $\pm$ 11.84 & 1.28 $\pm$ 0.09  & 3.11 $\pm$ 0.17 & 0.008 $\pm$ 0.003 & 0.42 $\pm$ 0.10  & 0.08 $\pm$ 325.54 &   & 1.0 \\ \hline
\end{tabular}
\end{table*}

\begin{table*}
\caption{Results of the broad band SED analysis on the sources at different epochs for the constant Bulk Lorentz factor $\Gamma$ = 15.}
\label{table-6}
\centering
\begin{tabular}{llrccccrc} \hline
      &       & Viewing      & Low energy     & High energy    & Eletron energy      & Magnetic      & Accretion &             \\
Name  & Epoch &  Angle      & particle index & particle index & density (cm$^{-3}$) & field (Gauss) & rate      & $\chi^2$/dof \\
\hline
PKS 1510$-$089 & A     & 2.97 $\pm$ 0.34    & 1.10 $\pm$ 0.57  & 4.24 $\pm$ 0.16  & 0.12 $\pm$ 0.04  & 0.84 $\pm$ 0.07  & 1.50 $\pm$ 0.31  & 0.8 \\ 
             & B      &  3.46 $\pm$ 0.24    & 1.12 $\pm$ 0.19  & 3.87 $\pm$ 0.12  & 0.19 $\pm$ 0.05  & 0.99 $\pm$ 0.06  & 1.25 $\pm$ 0.48  & 0.5 \\
             & C      &  4.17 $\pm$ 0.29    & 1.39 $\pm$ 0.23  & 3.55 $\pm$ 0.13  & 0.37 $\pm$ 0.09  & 0.64 $\pm$ 0.05  & 0.85 $\pm$ 0.22  & 0.7 \\
             & D     &  4.63 $\pm$ 0.70    & 1.10 $\pm$ 1.26  & 3.95 $\pm$ 0.18  & 0.21 $\pm$ 0.15  & 0.98 $\pm$ 0.16  & 1.26 $\pm$ 0.22  & 1.2 \\
             & E     &  3.41 $\pm$ 0.33    & 1.14 $\pm$ 0.27  & 3.63 $\pm$ 0.10  & 0.24 $\pm$ 0.08  & 0.68 $\pm$ 0.06  & 0.79 $\pm$ 0.28  & 1.3 \\
             & F      &  2.50 $\pm$ 0.25    & 1.10 $\pm$ 0.33  & 3.82 $\pm$ 0.12  & 0.07 $\pm$ 0.02  & 1.12 $\pm$ 0.06  & 1.50 $\pm$ 0.67  & 1.9 \\ \hline
             
3C 273       & A      &  3.17 $\pm$ 0.46 & 2.56 $\pm$ 0.27  & 4.00 $\pm$ 0.16  & 0.01 $\pm$ 0.003  & 2.21 $\pm$ 0.40  & 16.00 $\pm$ 3.69  & 1.9 \\
             & B      &  6.26 $\pm$ 0.29 & 1.65 $\pm$ 0.12  & 4.80 $\pm$ 0.40  & 0.11 $\pm$ 0.02  & 2.04 $\pm$ 0.16  & 16.00 $\pm$ 2.73  & 1.1 \\ \hline
3C 279       & A      &  4.14 $\pm$ 0.52    & 2.03 $\pm$ 0.46  & 4.24 $\pm$ 0.19  & 0.07 $\pm$ 0.02  & 1.08 $\pm$ 0.15  & 0.01 $\pm$ 0.22  & 1.7 \\
             & B      &  2.55 $\pm$ 0.28    & 1.36 $\pm$ 0.21  & 3.73 $\pm$ 0.11  & 0.03 $\pm$ 0.01  & 1.26 $\pm$ 0.09  & 0.01 $\pm$ 2.65  & 1.3 \\  
             & C      &  3.56 $\pm$ 0.22    & 1.24 $\pm$ 0.21  & 3.51 $\pm$ 0.09  & 0.22 $\pm$ 0.04  & 0.66 $\pm$ 0.04  & 0.01 $\pm$ 1.80  & 1.9 \\
             & D      & 2.68 $\pm$ 0.38    & 1.72 $\pm$ 0.26  & 3.70 $\pm$ 0.11  & 0.05 $\pm$ 0.01  & 1.52 $\pm$ 0.17  & 0.01 $\pm$ 5.90  & 3.1 \\ \hline
             
CTA 102     & A      & 3.71 $\pm$ 0.45 & 1.10 $\pm$ 0.70 & 4.37 $\pm$ 0.38 & 0.006 $\pm$ 0.003 & 0.32 $\pm$ 0.07  & 10.00 $\pm$ 5.86 & 1.4 \\
            & B      & 1.38 $\pm$ 0.49 & 1.25 $\pm$ 0.11 & 2.98 $\pm$ 0.18 & 0.009 $\pm$ 0.004 & 0.40 $\pm$ 0.12  & 0.25 $\pm$ 388.08 & 1.1 \\ \hline
             
\end{tabular}
\end{table*}

\begin{table}
\caption{Results of the linear least squares fit to the optical and $\gamma$-ray 
flux measurements, during different epochs for the sources PKS 1510-089, 3C 273, 
and 3C 279. Here R and P are the Spearman rank correlation coefficient and 
the probability for no correlation respectively.}
\label{table-7}
\addtolength{\tabcolsep}{-1.6pt}
\begin{tabular}{lcrrcc} \hline
Object         & Epoch  & Slope          & Intercept         & R    & P       \\ \hline
PKS 1510$-$089 & C   &   3.23 $\pm$ 1.07  &  27.11 $\pm$ 12.09 &   0.55 & 0.08 \\
               & E   &   1.45 $\pm$ 0.60  &  7.13  $\pm$ 6.73  &   0.57 & 0.07 \\
3C 273         & A   &  16.93 $\pm$ 3.52  & 156.64 $\pm$ 34.52 &   0.94 & 0.00 \\ 
3C 279         & B   &   4.78 $\pm$ 1.66  &  41.96 $\pm$ 17.90 &   0.50 & 0.06 \\
               & C   &   1.96 $\pm$ 1.45  &  12.02 $\pm$ 16.01 &   0.55 & 0.05 \\ \hline
\end{tabular}
\end{table}

\section{Summary}
We carried out detailed investigations of the correlation between optical and GeV 
flux variations in four FSRQs namely PKS 1510$-$089, 3C 273, 3C 279 and CTA 102. Our 
study includes (a) identification of epochs in those sources with anomalous optical-GeV 
flux variations, (b) analysis of the broad band SEDs of the sources in those 
epochs, (c) the analysis of $\gamma$-ray spectra and (d) analysis of optical-IR colour 
variations. The results of those analysis are summarized below:
\begin{enumerate}
\item All the four FSRQs studied here, namely PKS 1510$-$089, 3C 279, 3C 273 and 
CTA 102 showed varied correlations between optical and GeV flux variations. We 
found cases when (a) the optical and $\gamma$-rays are closely correlated, (b) epochs 
when there are optical flares without $\gamma$-ray and (c) epochs when 
there are $\gamma$-ray flares without optical counterparts. From our one zone leptonic model fit to the observed SED of all such epochs in the sources, we found that the regions giving rise to optical and $\gamma$-ray flares are co-spatial.

\item SED analysis indicates that the optical emission is often well explained by 
synchrotron emission process and the $\gamma$-ray emission is well explained by 
EC process with the seed photons from the torus. A Prominent accretion disk component is seen in the synchrotron part of the SEDs in PKS 1510$-$089, 3C 273 and the quiescent state SED of CTA 102. For 3C279, there is no evidence of an accretion disk component in the synchrotron part of the SED.
\item From model fits we found that (a) correlated optical and $\gamma$-ray flux variations are caused by increase in the bulk Lorentz factor (b) $\gamma$-ray flares with no optical counterparts are due to an increase in the bulk Lorentz factor and/or increase in the electron number density and (c) an optical flare with no $\gamma$-ray counterpart is due to an increase in the magnetic field.

\item The $\gamma$-ray spectra of the sources during various epochs are well represented by the LP model

\item Varied colour behaviours such as BWB trend and RWB trend are seen in our sample of sources.

\end{enumerate}

\section*{Acknowledgments}
We thank the referee for his/her comments that helped the authors
to improve the manuscript. This paper has made use of up-to-date SMARTS 
optical/near infra-red light curves that are available at 
www.astro.yale.edu/smarts/glast/home.php. Also, data from the Steward 
Observatory spectropolarimetric monitoring project were used. This program is 
supported by Fermi Guest Investigator grants NNX08AW56G, NNX09AU10G, NNX12AO93G, 
and NNX15AU81G. This research has made use of the High Performance Computing Facility of 
the Indian Institute of Astrophysics, Bangalore.

\section{Data Availability}
The multiwavelength data underlying this article are publicly available from the {\it Fermi-LAT{\footnote{https://fermi.gsfc.nasa.gov/ssc/data/access/}}, Swift-XRT and Swift-UVOT{\footnote{https://www.ssdc.asi.it/mmia/index.php?mission=swiftmastr}}, SMARTS{\footnote{http://www.astro.yale.edu/smarts/glast/home.php}} and Steward observatory{\footnote{http://james.as.arizona.edu/$\sim$psmith/Fermi/DATA/individual.html}}}.
\bibliographystyle{mnras}
\bibliography{example}




\bsp	
\label{lastpage}
\end{document}